\documentclass[12pt,preprint]{aastex}

\shorttitle{Absorption in Gamma Ray Burst afterglows}
\shortauthors{Stratta et al.}

\begin{document}

\title{Absorption in Gamma Ray Burst afterglows}

\author{G. Stratta\altaffilmark{1}, F. Fiore and L. A. Antonelli}
\affil{Osservatorio Astronomico di Roma, Via Frascati 33, I-00044 Monteporzio Catone, Italy}

\and

\author{L. Piro and M. De Pasquale\altaffilmark{1}}
\affil{Istituto di Astrofisica Spaziale \& Fisica Cosmica, C.N.R., Via 
Fosso del Cavaliere 100, I-00133, Roma, Italy}

\altaffiltext{1}{Dipt. di Fisica, Universita' ``La Sapienza'', P.le A. Moro 5, 
I-00187 Roma, Italy}

\begin{abstract}

We studied the X-ray and optical absorption properties 
of 13 Gamma Ray Burst afterglows observed by BeppoSAX.  
We found that X-ray absorption
in addition to the Galactic one along the line of sight is highly
statistically significant in the two cases with the best
statistics (probability $>99.9\%$).   
In three other cases the presence of X-ray absorption is 
marginally significant (probability $\sim 97\%$).
Measured rest frame absorbing column densities of hydrogen, $N_H$, 
range from 0.1 to 10.0 $\times10^{22}$ cm$^{-2}$ (at 90\% confidence level) 
assuming a solar metal abundance.  
X-ray absorption may be common, although the quality of present data
does not allow us to reach a firm conclusion.  
We found that the rest 
frame column densities derived from XMM and Chrandra data as quoted in the 
literature are in good agreement with the BeppoSAX estimated 
rest frame $N_H$ range, supporting our result.
For the same GRB afterglow sample we evaluated the rest frame 
visual extinction $A_{Vr}$. 
We fitted the optical-NIR afterglow photometry with a power law model 
corrected at short wavelengths by four different extinction curves. 
By comparing X-ray absorptions and optical extinction, we found that if a 
Galactic-like dust grain size distribution is assumed, a dust to gas ratio
lower than the one observed in the Galaxy is required by the data.
A dust to gas ratio $\sim$ 1/10 than the Galactic one, as in the 
Small Magellanic Cloud (SMC) environment, has been tested using 
the SMC extinction curve, which produces good agreement between the 
best fit $N_H$ and $A_{Vr}$. We note, however, that the best fit 
$N_H$ values have been obtained by assuming solar metal abundances, 
while the metallicity of the SMC ISM is $\sim$ 1/8 the solar one 
(Pei 1992). If such low metallicity were assumed, the best fit $N_H$ values 
would be higher by a factor of $\sim7$, providing a significant
increase of the $\chi^2$.
Alternative scenarios to explain simultaneously the optical and X-ray
data involve dust with grain size distributions biased toward 
large grains. Possible mechanisms that can bring to such a grain size
distribution are discussed.

\end{abstract}

%\keywords{gamma rays: bursts, X-ray -- absorption: dust extinction}

\section{Introduction}

Soon after the $\gamma$-ray flash, optical and X-ray afterglows of
Gamma Ray Bursts (GRB) are among the brightest sources in the sky at
cosmological redshifts.  More than thirty GRB redshifts have been measured
to date and their distribution ranges from 0.168 (Greiner et al. 2003) 
to 4.5 (Andersen et al. 2000) with a median of z $\sim 1.1$ (excluding GRB 980425 if at z=0.0085). 
Follow-up observations of the GRB localized by
BeppoSAX, the IPN and by the RXTE and HETE2 satellites,
show that tens of minutes after the GRB the optical afterglow can 
be as bright as R=14-16 mag; a few hours later it can still be as bright 
as R=17-19 mag. An exceptionally bright example is the case of 
GRB 030329 for which the optical afterglow 
reached R=12.7 mag at 1.5 hours from the GRB and it decreased down to R=19 mag 
after $\sim 10$ days (e.g. Price et al. 2003; Stanek et al. 2003).
The X-rays afterglow can be as bright as the Crab Nebula a few minutes
after the GRB, while 5-8 hours later it can be as bright as a bright
AGN, i.e. $\approx$mCrabs (see Frontera et al. 2000, also Fiore et al. 2000
for an estimate of the GRB afterglows logN-logF in the range
0.5-2.0 keV, where fluxes are integrated from minutes up to 
hours after the GRB event). This opens up the
perspective for gathering spectra of high quality of sources at
cosmological redshifts, provided that the afterglow can be observed in
such short time scales.  Spectral studies of GRB afterglows can provide
crucial data on the environments in which GRB occurs. This can give us
both important constraints on the nature of the GRB progenitor and on
the interstellar matter (ISM) in the GRB host galaxies.

Candidate GRB progenitor include: 'collapsars' (Woosley 1993, Paczynski 1998; 
MacFadyen \& Woosley 1999), 'supranovae' (Vietri \& Stella 1998), 
mergers of two neutron stars or a neutron star and a black hole 
(e.g. Eichler et al. 1989; Paczynski 1990; Narayan, Paczynski \& Piran 1992; 
Meszaros \& Rees 1992).  

Multiwavelength afterglow observations suggest the 
association of ``long GRB'' (for which the duration 
of the prompt event peaks at $\sim20$s, e.g. Norris, Scargle \& Bonnell 2000) 
progenitor with massive stars. 
Some indications are, among others: i) the GRB locations relative 
to their host galaxy (Bloom, Kulkarni \& Djorgovski 2002a); ii) 
the detection of iron lines in the X-ray 
afterglow spectrum of five GRB (Piro et al. 1999, 2000; Yoshida et al. 1999; 
Antonelli et al. 2000; Amati et al. 2000); iii) the presence of late
re-brightening in the light curves of some afterglows, interpreted as 
the evidence of an underlying supernova (e.g. Bloom et al. 2002b);
 iv) the presence of supernova features emerging in the optical afterglow 
spectra of GRB 030329 (Stanek et al. 2003, Hjorth et al. 2003; 
Kawabata et al. 2003) and of GRB 021211 (Della Valle et al. 2003).
Because of the short lifetimes of massive stars, it is likely that their 
collapse happens close to their star forming region, that is, close to a 
dense and dusty environment (e.g. Fryer, Woosley \& Hartman 1999; 
Perna \& Belczynski 2002a). 
Therefore, one important element in favor of massive star progenitor 
would be the evidence of strong dust extinction.
Indeed, the non detection of the $\sim 60\%$ of X-ray afterglows 
at optical wavelengths (dark GRB), favors this hypothesis.
However, the optical spectra of GRB afterglows, 
 do not generally show strong reddening (Simon et al. 2001; 
Galama \& Wijers 2001).
Moreover, Lazzati, Covino \& Ghisellini (2002)
show that even if dark GRB are located in the innermost regions of 
Galactic-like molecular clouds, in several cases the corresponding 
visual extinction is not enough to hide their optical afterglows.
Therefore, a more accurate study of the GRB environment is needed.

Prompt observations of GRB afterglows offer a new and distinctive path
for the study of the matter in the immediate surroundings of the GRB, 
r$\sim$1-10~pc (Perna \& Loeb 1998; B\"ottcher et al. 1999; Perna et al. 2002b;
Perna \& Lazzati 2002c; Fox et al. 2003) 
and in the GRB host galaxy (Ciardi \& Loeb 2000; Fiore 2001; 
Bloom et al. 2002a; Savaglio, Fall \& Fiore 2003) 
X-ray and optical-UV spectroscopy can tell us about gas density and
ionization status, metal abundances, dust content and kinematics of
the matter in the GRB environment.  This can be done by using both
emission features and absorption features 
(see e.g. Kumar \& Narayan 2003; B\"ottcher et al. 1999; 
Piro et al. 2000; Ghisellini et al. 2002).
Although absorption features are more difficult to
study than emission lines, they carry with them unbeatable
information, because they probe matter along a single beam, i.e. along
the line of sight to the background beacon.  This greatly reduces the
complications related to the matter geometry and dynamics, which
strongly affects emission line studies. Even when such emission or 
absorption features are not detected or are not resolved, 
X-ray spectra can provide useful information from measures of low
energy cut-offs due to photoelectric absorption

Moreover, GRB afterglows can be used to study the ISM of their host galaxies.
High redshifts galaxies have so far been studied mainly via
`Lyman-break galaxies' (e.g. Steidel et al. 1999), and via galaxies
which happen to lie along the line of sight to bright quasars, notably
the `Damped Lyman-alpha' systems (DLA, e.g. Pettini et al. 1997, 1999).  
However, probably neither of these methods gives an
undistorted view of the bulk of the typical high redshifts galaxy.
Lyman-break galaxies are characterized by pronounced star-formation
and their inferred chemical abundances may be related to these regions
rather than being representative of typical high z galaxies (see
e.g. Steidel et al. 1999).  DLA studies are likely to be biased
against dusty systems, since these would hide the background quasars
in color-based surveys. Furthermore, DLA are often identified with
dwarf or Low Surface Brightness galaxies and therefore may not be
representative of the full high z galaxy population.  On the other
hand, GRB seem to occur well within the main body of their host
galaxies, not in the outer halos (Bloom et al. 2002a), 
therefore, GRB afterglows can provide a powerful, 
independent tool to study the ISM
of high redshifts galaxies (Castro et al. 2003, Mirabal et al. 2002, 
Fiore 2001; Savaglio et al. 2003). 
The latter authors compared metal column density in GRB host
galaxies with those of DLAs. They find that while the column densities
of iron and magnesium are similar or slightly higher than that in DLAs,
the column of zinc are significantly higher.  Since iron and magnesium,
 unlike zinc, tend to be depleted in dust, this finding indicates 
that a significant fraction of these elements is in dust since not 
observed through absorption lines, thus GRB are probably probing 
denser and dustier regions, than DLAs.

We present in this paper the results of a systematic spectral analysis
of 13 GRB afterglows observed by BeppoSAX in X-rays, and by several
ground based telescopes in optical-UV, aimed at measuring and
constraining the X-ray absorption along the line of sight as well as
the extinction in the optical-UV bands.  The comparison of the results
obtained in the two spectral band will provide constraints on the dust
to gas ratio of the absorbing matter and on the dust properties, as
seen from several hours to a few days from the GRB event.

The paper is organized as follows: \S 2 shows the BeppoSAX 
X-ray afterglow selected sample; \S 3 presents the X-ray data reduction 
and spectra extraction procedures, along with a discussion on the systematics 
affecting the column density measurements;  
\S 4 presents the optical photometry of 9 GRB afterglows 
of our sample with a detected optical counterpart;
results are presented and discussed in \S 5 and \S 6 respectively.

\section{The BeppoSAX X-ray bright afterglow sample} 

We selected from the BeppoSAX archive all
the X-ray afterglows observed during a period of four years 
(from February 1997 to February 2001) that have enough counts 
in the Medium Energy Concentrator Spectrometer (MECS) to extract a 
spectrum and to perform a spectral analysis,
that is, with a signal to noise ratio $S/N\ga7$ 
in the 2.0-10.0 keV. We also required that the 
Low Energy Concentrator Spectrometer (LECS) 0.1-2 keV
spectra contain at least 20 counts, or that 
the counts expected in this LECS band,  based on 
simulations performed using the best fit power law model 
obtained from the MECS spectrum and assuming a
column density equal to the Galactic value along the line of sight,
are also smaller than the above threshold 
at the $2\sigma$ confidence level, not to miss
bright and highly obscured afterglows.
The number of afterglow observations satisfying these requirements is 13.  

Table \ref{tbl-1} summarize the selected GRB X-ray afterglows along with the Galactic hydrogen column density along the 
line of sight to each GRB, as determined 
from the coarse, $\sim3$ deg resolution, maps of Dickey \& Lockmann (1990), 
the host galaxy redshifts if known, the R.A. and Dec. coordinates and the optical detection. Redshifts measurements are from optical 
spectroscopy except for GRB 000214 (Antonelli et al. 2000), for which a 
redshifts of $0.47\pm0.06$ was inferred assuming that the observed
emission line is the iron $K_{\alpha}$ line at 6.7 keV (rest frame).

\section{BeppoSAX data reduction and analysis}

The GRB afterglows observations were performed with the BeppoSAX
Narrow Field Instruments (NFI), LECS 0.1-10 keV (Parmar et al. 1997), MECS
1.3-10 keV (Boella et al. 1997), HPGSPC 4-60 keV (Manzo et al. 1997)
 and PDS 13-200 keV (Frontera et al. 1997).  LECS (1 unit) and MECS (3 units)
are imaging gas scintillation proportional counters, the HPGSPC is a
collimated High Pressure Gas Scintillation Proportional Counter and
the PDS consists of four collimated phoswich units.  We report here
the analysis of only the LECS and MECS data because GRB afterglows
are usually many times fainter than the PDS and HPGSPC internal
backgrounds.

After 1997 May 6$^{\it th}$ MECS observations were performed with
MECS2 and MECS3 units only because on this date a technical failure caused the
switch off of the MECS1 unit. For all the selected GRB sample, data from 
the MECS2 and MECS3 units were combined 
together, after gain equalization, to increase the signal to noise ratio.  
In the case of GRB 970228 (that is the only GRB in our sample observed 
before 1997 May) we used the data from MECS3 only, since the source
happened to lie just below the Beryllium strongback support of the
MECS1 and MECS2 windows (which basically absorbs all photons below
about 5 keV).

Standard data reduction was performed using the SAXDAS software
package version 2.0 following Fiore, Guainazzi \& Grandi (1999).  

LECS and MECS spectra were extracted from regions of radii between 3
and 8 arcmin, chosen to maximize the count rate in the 0.5-2 keV and
2-10 keV band respectively.
The choice of the extraction radius depends on the 
source and background intensity, and on the instrument Point Spread
Function, PSF. The 
90\% encircled energy generally
is within 3-4 arcmin for the MECS and within 
6-8 arcmin for the LECS (Fiore et al. 1999).

Background spectra were extracted in detector coordinates from high
Galactic latitude ($|b|>20$ deg) ``blank fields'' 
(averaged LECS and MECS pointing of sky regions without 
strong X-ray sources, for a total exposure of about half a million 
of seconds in both instruments, Fiore et al. 1999)
using the same source extraction regions. 
``Local'' background cannot be safely used when analyzing LECS and 
MECS spectra, because of the relatively broad PSF and because 
the background strongly varies with the position in the
detector (Fiore et al. 1999 and Chiappetti et al. 1998).  We have
compared the mean level of the background in the LECS and MECS ``blank
fields'' observations to the mean level of the background in the GRB
observations using source free regions at various positions in the
detectors. The ``local'' MECS background count rate is within 2\% than
that in the ``blank fields''. On the other hand, the ``local'' LECS
background count rate differs from that of the ``blank fields'' by up
to 20\% in 7 cases. In these cases the ``blank fields'' background
spectrum was multiplied by appropriate correction factors before
subtraction from the source spectrum.

Usually, the position of GRB afterglows is not know with 
accuracies better than several arcminutes at the moment of the 
BeppoSAX NFI pointing.
On these scales the vignetting of the telescopes varies significantly
(see e.g. Fiore et al. 1999).  Furthermore, the LECS thin
polypropylene window is supported by a mesh of grids of tungsten wires
that introduces a complicated pattern of obscuration. The actual value
of obscuration strongly depends on the exact position of the source
centroid in detector coordinates.  The spacing of the thicker wire
grid is of about 2 arcmin, much less than the uncertainty on the
afterglow position. This means that the afterglow position in detector
coordinates with respect to the thick wire grid cannot be known `a
priori' as well as the level of the grid obscuration. For these two
reasons we therefore not used the standard on axis effective area
files, but we evaluated the instrument effective area at the afterglow
position in detector coordinates 
(we used the SAX/LECS data analysis software {\sc lemat}).  
These `ad hoc' effective area file
takes correctly into account the energy dependent telescope vignetting
and the photons absorbed by the LECS support mesh wires.  On axis
redistribution matrices, from the December 1999 release have been used
for both LECS and MECS, since these depend very little on the source
position in the detectors.

Table \ref{tbl-2} gives the LECS and MECS exposure times,
optimum extraction radius, background and source count rates.  
Note that LECS exposure times are usually smaller than MECS
ones since it is operated during dark time only.

Table \ref{tbl-3} gives the MECS off-axis angles,
the time interval between the GRB trigger and the NFI pointing, 
the LECS `raw' detector pixel coordinates of the source, 
at which the effective area has been computed.

To get Gaussian statistics and so to ensure the applicability of the 
$\chi^2$ test to evaluate the goodness of our fits, LECS and MECS 
spectra were rebinned to obtain at 
least 20 counts per energy channel.
We then performed simultaneous LECS and MECS spectral fitting using the bands
0.1(or 0.4 keV, see below) 4.0 keV and 1.6-10.0 keV in the two
instruments respectively (see \S 5.1). 
Errors quoted in this paper
for the column density and the spectral index are 90\% confidence intervals
for two parameters of interest ($\Delta\chi^2=4.61$), 
unless differently specified.

\subsection{Systematic errors on  Column Density estimates}

\subsubsection{Galactic Column Densities}

The knowledge of the Galactic column density along the line of sight
to each burst is crucial for our study.  Relatively small errors in
the evaluation of the Galactic columns would translate in a larger
errors on the rest frame absorption, since best fit $N_H$ values scale
with redshift as $(1+z)^{\sim2.6}$ (the photoelectric cross section
scaling with energy roughly as $E^{-2.6}$).  The values in Table \ref{tbl-1} 
have been obtained interpolating on a grid with spacing of
3 degrees (Dickey \& Lockmann 1990). It is well known that the Galactic column density can vary
on much smaller scales (Elvis, Lockman \& Wilkes 1989). To evaluate the average
magnitude of this variation we compared the Galactic column density
obtained using the coarse 3 degrees 21 cm emission maps along the line
of sights of a sample of AGN for which an accurate 21 cm emission
measurements were already performed by using a 21 arcmin beam NRAO 140
ft telescope at Green Bank by Elvis et al. (1989).  We found that 
the average ratio between the two determinations is close to 1 and
that the standard deviation is of 0.16, independent of the value of
the Galactic column density.  The highest positive deviation found is
of 50\%.  We therefore performed for each GRB three series of fits,
one with the nominal Galactic $N_H$ value  
and the other two by increasing this value by 16\% and 50\% respectively, to
check whether additional extra galactic absorption would still be
significantly even in case the Galactic column has been underestimated
in the Dickey \& Lockmann (1990) maps.

Another tracer of the column density of matter in a given direction is
the infrared emission at $100\micron$. This can therefore be used as a
further check of the Galactic column density values given in 
Table \ref{tbl-2}.  
We used the IRAS faint source survey to search for any
``cirrus'' contamination along the line of sight of each GRB. We used
a 10 to 30 arcmin radius search region to measure the $100\micron$
MJy/sr brightness $S_{100}$. By adopting $dS_{100}/dN_H=0.5-2.0$
MJy/sr 10$^{20}$ cm$^{-2}$ as typical Galactic dust to gas ratio
(Reach, Heiles \& Koo 1993; Heiles, Reach \& Kooh 1988; de Vries,
Heithausen \& Thaddeus 1987) we estimated the implied $N_{H Gal}$
values and then compared them with the maps of Dickey \& Lockmann
(1990). All the inferred $N_{H Gal}$ values reveal no absorption excess
due to presence of cirrus along the analyzed line of sight.

\subsubsection{Background subtraction}

The background subtraction procedure previously described works well 
if the Galactic column density along the line of sight to the GRB, is not
much higher (or lower) than that in the ``blank fields''. Otherwise
the background low energy spectrum (e.g. below 0.3 keV) would be very
different from that of the ``blank fields''. Blank fields are all at
high Galactic latitude, and the average Galactic column density along
their line of sight is $\sim3\times10^{20}$ cm$^{-2}$.  Several
afterglows in our sample have a Galactic column density much higher than this
value (see Table \ref{tbl-2}). 
Adopting the ``blank field'' background spectrum 
for those afterglows at Galactic $N_H\ga5\times10^{20}$cm$^{-2}$
it would results in an over-subtraction of the background
below 0.3-0.4 keV and in a consequent overestimate of the intrinsic
absorbing column.  
As an example, the 0.2 keV flux would decrease by $\sim 90\%$
by increasing $N_H$ from $3\times10^{20}$cm$^{-2}$  to 
$5\times10^{20}$cm$^{-2}$. At 0.4-0.5 keV the flux would decrease 
by much smaller amount, of the order of 10-20\%. We therefore decided to 
limit the spectral analysis of the afterglows with Galactic 
column densities higher than
$5\times10^{20}$ cm$^{-2}$ to energies higher than 0.4 keV.

\subsubsection{LECS-MECS calibration of the normalization factor}

A Constant normalization factor must be introduced in the fitting models in
order to take into account the systematic error in the knowledge of
the absolute LECS and MECS normalization factor (Fiore et al. 1999).
Assuming the MECS as reference instrument, the expected factor between
LECS and MECS for constant sources is about $0.9\pm0.2$.
However, GRB afterglows are not constant sources: their flux decreases
usually as a power law with index $\delta$ in the range $-1\div -1.5$
(see e.g. Costa et al. 1999; Piro 2002a).   Moreover, the non
completely simultaneous observations of the LECS and MECS instruments
due to the switch-off of the LECS during part of the orbit, introduces
a further complication in the determination of the normalization factor. 
We have therefore decided to leave completely free the normalization
factor in the fits. 

We are aware that letting the normalization factor be free to vary 
introduces an additional degree of freedom in the fit and 
consequently increases the uncertainty on the derived $N_H$.
In fact, a truly curved spectral shape toward the lower 
energies could be mimicked, in spectra of low statistics, by a lower 
MECS to LECS normalization factor. 
As an example a factor of $\sim2$ lower normalization factor, 
could mimic a factor of $\sim3-4$ lower $N_H$.
The $68\%$ LECS-MECS normalization factor confidence intervals 
are reported in Tables \ref{tbl-5} and \ref{tbl-6}.

\section{Optical data}

For 9 of the 13 selected GRB with X-ray afterglow an optical 
afterglow was also discovered (see Table \ref{tbl-1}).  
We collected the optical and near infrared magnitudes 
of these nine GRB from the literature
to study the extinction properties through the comparison 
of dust extinction models to the afterglow
Spectral Energy Distribution (SED).

We selected, among the large number of papers reporting
photometric data on a given GRB optical afterglow, 
those works containing the widest possible frequency coverage and
the best and most complete documentation. We list and discuss 
the references used for each GRB afterglow in the 
appendix section.

Observations in different bands are usually performed at different
times. Since optical afterglows are sources with fluxes rapidly 
decaying in time, their magnitudes have been extrapolated to a single 
specific time $t_0$.
The $t_0$ was selected to be as close as possible to the 
first optical observations, to keep the statistical uncertainties on 
the observed magnitudes small. 
The extrapolation of the observations at 
different bands at $t_0$ was performed by adopting the best fit
decay index quoted in the literature. 

Corrections for the Galactic interstellar extinction have been
performed, unless explicitly specified, using the 
IRAS 100 $\micron$ E(B-V) maps by Schlegel,
Finkbeiner \& Davis (1998, SFD) and we derived the extinction at different
wavelengths using the extinction curve
parameterization by Cardelli, Clayton \& Mathis (1989) assuming
$R_V=A_V/E(B-V)=3.1$.

Table \ref{tbl-4} gives for each GRB the magnitude in each photometric bands, 
extrapolated at $t_0$ (see notes on Table \ref{tbl-4}) 
and corrected for Galactic extinction. We conservatively 
kept the error on the magnitudes to 0.03 mags when the published error 
is smaller that this value.

GRB 000926 is not part of our BeppoSAX sample because its 
X-ray data has poor statistics (see section 2).
However, it has relatively high quality Chandra data and a complete
optical-NIR photometry. Furthermore, GRB 000926 has been observed
at a medium-high resolution with Keck ESI (Castro et al. 2003) and 
Savaglio et al. (2003) were able to derive information on the metal 
column densities from this spectrum. 
For these reasons, we decided to include it in our analysis.
We take the photometry of GRB 000926 (U, B, V, R, I, J, H and K) from Fynbo 
et al. (2001) and the Galactic extinction in the direction of
this burst from the SFD maps as $A_V=0.07$. The redshift of GRB 000926 is
$z=2.0379\pm0.0008$ (Castro et al. 2003).

The optical to near infrared SEDs of the
nine GRB afterglows are shown in Figure \ref{f1}.

\section{Results}

In this section we summarize the results of afterglow spectral
fitting, performed separately in the X-ray and in the 
optical-NIR energy ranges. Our goal is to estimate 
the absorption in the soft X-ray energy 
and extinction in the optical band.

\subsection{X-ray spectral fittings}

At the time of the NFI observations (on average after $\sim 10$ hours
after the GRB event, see Table \ref{tbl-3}), 
afterglow X-ray spectra are usually well described by simple power laws.
We therefore performed a spectral fit by adopting a simple power law 
model with two photoelectric absorption component: one representing the
absorption from our Galaxy and the other representing the 
additional absorption along the line of sight. 

The quality of the data does not allow to fit for the absorber redshift.
We therefore present in Table \ref{tbl-5} (0.1-10 keV
band) and Table \ref{tbl-6} (0.4-10 keV band) best fit parameters obtained with the additional absorber at the redshift of the GRB, or at z=1 when the redshift of the GRB is unknown, since GRB redshift distribution
peaks at about this value (Djorgovski et al. 2001).

Tables \ref{tbl-5} and \ref{tbl-6} also give the results of the fits 
performed fixing the
redshift of the additional absorbing column to zero.  Figure
\ref{f2} shows the best fit additional column density at z=0
against the Galactic column. No trend is evident, supporting the
robustness of our analysis.

Figure \ref{f3} shows the best fit additional absorbing column density 
 measured at z=0 and at z=z$_{GRB}$ (or z=1 for the GRB with unknown 
redshift) as a function of the redshift. 
Since the photoelectric cross section scales with energy roughly 
as $E^{-2.6}$, effective $N_H$ values scale with redshift as 
$(1+z)^{2.6}$. Consequently, fitting the absorption in the 
GRB frame the resulting column is highly magnified with respect 
to that fitted at z=0 and small errors in the evaluation of the 
Galactic column would translate in a strong 
trend of the additional column with z toward high $N_H$ values 
at high z. This is not observed, supporting the robustness of 
our analysis.

To evaluate the statistical significance of a given additional $N_H$
measurement we used the F test (Bevington \& Robinson 1992).  
We found that in two cases,
 GRB 990123 and GRB 010222, the improvement of the 
$\chi^2$ including the additional absorber to the fit is highly 
significant (Figure \ref{f4}), 
probability $>$ 99.99\%.  
In both cases the probability remains higher
than 99.99\% increasing $N_{H Gal.}$ by 16\% (that correspond to
 68 \% of uncertainty on our estimate of the Galactic column along 
the line of sight, see \S 3.1.1). 
Even increasing $N_{H Gal.}$ by 50\% (the highest
deviation we expect from our estimates of $N_{H Gal.}$) the
probability remains higher than 99.99\% for GRB 010222 and higher than
98 \% for GRB 990123. 

In three other cases (GRB 980703, GRB 990510 and GRB 001109) the improvement
of the $\chi^2$ including the additional absorber to the fit is
marginally significant (probability of 97\%).  In all these cases
increasing $N_{H Gal.}$ by 16\% reduces the statistical significance
of the $\chi^2$ improvement below the threshold of 95\%.

We conclude that the detection of absorption in addition to
the Galactic one is robust against statistical and systematic
uncertainties in  2 of the 13 GRB analyzed (GRB 990123 and
GRB 010222).  
We note that the spectra of GRB 990123 and
GRB 010222 are also those with the best statistics. 
However, fitting the spectra of  two other GRB afterglows 
with lower but still reasonably good statistics, 
GRB 971214 and GRB 970228, including an additional absorber, does 
not produce a significant improvement in $\chi^2$ (probability of 
$93.3\%$ and $90.8\%$ respectively).
For the rest of the sample, additional absorption may be common, 
although the quality of the data in the majority of the cases does 
not allow highly significant detection.

Assuming that the most of the matter responsible of absorption is
at the GRB redshift, we found $N_H$ rest frame values in the range 
$0.1 - 10.0 \times10^{22}$ cm$^{-2}$ (90\% confidence level), 
similar to the column usually observed toward the disk and 
the bulge of our Galaxy (see Figure \ref{f3}). 

\subsubsection{GRB 010222}

A more detailed analysis has been performed for this burst which is 
the second brightest X-ray afterglow of the selected sample and
shows a peculiar behavior at low energies (Figure \ref{f4}).

We note from Table \ref{tbl-5} that our $N_H$ measurement for this burst, 
is somehow lower, but still consistent with the $N_H$ reported by
in't Zand et al. (2001) ($N_H=2.5\times 10^{22}$ cm$^{-2}$) using the same 
BeppoSAX data (0.3-10.0 keV). 
On the other hand, our measure is higher than that one reported
by Bj\"ornsson et al. (2002) $N_H=(6.5\pm0.11)\times 10^{21}$cm$^{-2}$, 
using Chandra observations (0.5-10.0 keV). 
We note, however, that the Chandra data are affected by pile-up. 
This introduces an additional systematic uncertainty in the
spectral parameter estimates.

The low energy absorption for this burst, may show possible 
complexities and/or variability.  
In fact, the presence of a neutral absorber produces a good fit at 
energies $\ga0.5$ keV but at lower energies an excess 
with respect to the best fit model is observed (Figure \ref{f5}).
There are at least two possible explanations for these residuals: 
1) the recovery of the spectrum below the energies of strong 
absorption edges of OVII (0.74/(1+z)=0.30 keV), 
OVIII (0.87/(1+z)=0.35 keV); 
2) the spilling of the emission spectrum through a leaky absorber.
We tested both models. 
First we fitted to the data a ionized absorber
model (`absori' in XSPEC) at the redshift of the host galaxy.
We obtained a good fit ($\chi^2=92.0$ with 110 degrees of freedom, d.o.f.) 
with a rest frame column density of $1.8^{+1.2}_{-0.9}\times10^{22}$ cm$^{-2}$ 
and a ionization parameter of $\xi=7.6^{+17.4}_{-7.1}$ (Figure \ref{f5}).
The probability for a chance improvement in $\chi^2$ is 0.3\% 
using the F-test. 
Second, we fitted a partial covering absorber (`zpcfabs' in XSPEC). 
In this case we obtain a similarly good fit ($\chi^2=92.5$ with 110 d.o.f.), 
a rest frame column density of $1.9^{+1.3}_{-0.9}\times10^{22}$ cm$^{-2}$
and a cover fraction of $0.8\pm0.1$. The probability for a chance  
improvement in $\chi^2$ is 0.4\% using the F-test.

To check for the presence of a narrow iron K$_\alpha$ emission line
we added to the simple power law model a Gaussian component,
fixing its width to $\sigma=0.01$ keV and letting its energy to be
free to vary in the range 6.4-6.9 keV rest frame. 
We found an upper limit to the line intensity 
of $I=1.7\pm10^{-5}$ photons s$^{-1}$ 
cm$^{-2}$ (90\% confidence level) and an upper limit to its equivalent 
width (EW) of 0.15 keV. 

\subsubsection{Additional X-ray observations}

We compared our BeppoSAX results with
other rest frame column density measures that became available, 
thanks to Chandra and XMM data for 9 additional GRB at known redshift.

Ballantyne et al. (2002) found a 90\% upper limit of
$N_H=0.64\times10^{22}$ cm$^{-2}$
 at z=1.0 for GRB 991216 using 0.4-8.0 keV Chandra data. 

Piro et al. (2001) and (2002b), estimated for GRB 000926 and for GRB 000210,  
$N_H=4.0^{+3.5}_{-2.5}\times10^{21}$ cm$^{-2}$ at 
 z=2.04 and $N_H=(5\pm1)\times10^{21}$ cm$^{-2}$ at z=0.846 
respectively, using Chandra {\it and} BeppoSAX data.
Note here that Harrison et al. (2001) do not present evidences 
for significant absorption in addition to the Galactic one 
for GRB 000926 at the redshift of the host galaxy using Chandra data.

Reeves et al. (2002) and Borozdin \& Trudolyubov (2003) do not 
detected any significant absorption above the Galactic value 
in the X-ray spectra of GRB 011211 using XMM data. 

Watson et al. (2002) measured a rest frame 
$N_H=(1.3\pm0.2)\times10^{22}$cm$^{-2}$ at $z=1.8$ 
for GRB 020322 using XMM. 
However, the uncertainty on the redshift estimated directly from 
the X-ray data ($z=1.8^{+1.0}_{-1.1}$) is large.
The confidence interval plotted for this burst in 
Figure \ref{f3} includes the uncertainty over z in addition to the statistical
 uncertainty.

Mirabal et al. (2002b) measured $N_H=(4.7\pm3.7)\times10^{21}$ cm$^{-2}$ 
for GRB 020405 at z=0.690, using Chandra data.

Butler et al. (2003) found no additional absorption for the
X-ray afterglow spectra of GRB 020813 and GRB 021004 using Chandra data.
Holland et al. (2002) found no absorption
in the vicinity of GRB 021004 (z=2.33) with an upper limit of 
$N_H=3.4\times10^{21}$cm$^{-2}$ using Chandra data.

Mereghetti et al. (2003) found $N_H=6.8^{+1.8}_{-3.8}\times10^{22}$ cm$^{-2}$
at $z=3.9\pm0.3$ for GRB 030227 using XMM data. 
On the other hand, the redshift estimated using the X-ray emission lines
is $z=1.39^{+0.03}_{-0.06}$ (Watson et al. 2003). 
Considering a possible blueshift from the X-ray emitting 
plasma, flowing at a velocity of $\sim$0.1c 
(Reeves et al. 2002; Butler et al. 2003), the progenitor
redshift should be at $z\sim 1.6$. The rest frame column at this redshift
would be $N_H=1.3^{+0.3}_{-0.7}\times10^{22}$ cm$^{-2}$.

All these values are in agreement with our rest frame $N_H$ range. 
We plot the above $N_H$ determinations along with our BeppoSAX 
measurements in Figure \ref{f3}.

\subsection{Optical-NIR spectral fittings}

According to the standard fireball model (see Sari, Piran \& Narayan 1998) 
at $>$1-2 days from the GRB event, the electron population accelerated by
shock mechanisms in a power law energy distribution, is expected to
radiate in the so called slow cooling regime. 
In this regime, if the optical range $\nu_o$ is at frequencies higher 
than the peak of the emission, the spectrum can be well described by a 
broken power law with indices $\alpha=(p-1)/2$ for $\nu_o<\nu_c$ 
and $\alpha=p/2$ for $\nu_o>\nu_c$, where $\nu_c$ is the cooling 
frequency and $p$ is the electron energy distribution index.
To evaluate the rest frame visual extinction $A_{Vr}$ we 
fitted the optical to near infrared data with the following 
model:
$$f(\nu_o)=C\nu_o^{-\alpha} e^{-A(\nu_o(1+z))}
\eqno(1)$$

where C is the spectral normalization constant, $A(\nu_o(1+z))$ is the rest 
frame extinction and z is the GRB redshift.

Since the optical-NIR photometry covers a rather narrow frequency range 
and the number of independent photometric points for each afterglow 
is usually small, we could not obtain from the fits strong 
constraints on {\it both} the emission spectrum and the rest frame 
extinction. 

We estimated the electronic spectral index $p$ 
from each X-ray afterglow spectral index $\alpha_X=p/2$ (Tables \ref{tbl-5} 
and \ref{tbl-6}), assuming that the X-ray frequencies $\nu_X$ are 
above the cooling frequency $\nu_c$ (Berger, Kulkarni \& Frail 2003).
On the basis of the $p$ measures, we then estimated the
optical spectral index according to the fireball model (Sari et al. 1998) 
considering two possible cases, where
the optical energy range is below or above the cooling frequency,
respectively $\alpha_1=(p-1)/2$ for $\nu_o<\nu_c$ or $\alpha_2=p/2$ 
for $\nu_o>\nu_c$, or $b_1$, $b_2$ if the flux is expressed in wavelength 
(see Tables \ref{tbl-7} and \ref{tbl-8}). We have then estimated the 
extinction uncertainties taking into account the 
uncertainties on the estimated optical spectral index $b_1$ and $b_2$.

For GRB 000926, that is not included in our X-ray selected
sample, we took the electronic spectral index derived 
from broad band modeling results published in previous works.
The interpretation of the broad band spectrum for this burst is not
univoque. However, similar values are obtained for the $p$ parameter.
Piro et al. (2001) find a best fit model with $\nu_X<\nu_c$, from which
the authors derive $p=2.6\pm0.3$. 
Harrison et al. (2001) adopt a different continuum 
model with $\nu_c<\nu_o<\nu_X$ and they find a minimum 
$\chi^2$ for $p=2.43\pm0.06$.  
Due to the uncertainties on the cooling frequency energy regime, 
we tested for this afterglow both the possible cases, 
$\nu_o<\nu_c$ and $\nu_o>\nu_c$, assuming $p=2.6\pm0.3$.

The model in equation (1) has been convolved with the 
transmission functions of the optical and near infrared filters and
then converted into a magnitude using the photometric zero points 
from Fukugita, Shimasaku \& Ichikawa (1995) for the optical range and 
from Bersanelli, Bouchet \& Falomo (1991) for the NIR range.
We then compared our results with the observed optical-NIR afterglow SEDs
using a $\chi^2$ minimization technique.
 
Four different extinction curves have been tested and best fit
rest frame visual extinction $A_{Vr}$ have been estimated.  We used: 
{\it i)} the Galactic extinction curve from Cardelli et al. (1989) 
(``G'') assuming $R_V=A_V/E(B-V)=3.1$; {\it ii)} the 
extinction curve of the Small Magellanic Clouds 
(``SMC'') from Pei (1992); {\it iii)} the extinction curve computed by 
Calzetti, Kinney \& Storchi-Bergmann (1994) and Calzetti (1997) for a 
sample of local starburst galaxies (``C''); 
{\it iv)} the extinction curve computed from simulations by 
Maiolino, Marconi \& Oliva (2001a) based on a dust model skewed toward
large grains. 
The Galactic extinction curve is reasonably well described
by a dust model were grain sizes are distributed as a power law, namely 
$dn(a)\propto a^{q}da$, where $n(a)$ is 
the number density of grains with size $\la a$, $q=-3.5$ and  
$a_{min}=0.005\mu$m $< a < a_{max}=0.25\mu$m 
(Mathis, Rumpl \& Nordsieck 1977). 
The Maiolino et al. (2001a) extinction curve Q1 used the same
index $q=-3.5$, but a higher maximum grain size $a_{max}=10\mu$m.  

All these extinction models are plotted in Figure \ref{f6}.

Note that the Q1 and the C extinction curves are
flatter than the rough $\approx \frac{1}{\lambda}$ behavior of the 
Galactic extinction curve. This gives rise to a more gray extinction.
Note also that in the C and in the SMC curves, 
the 2175 \AA\ hhump is absent, while
is quite prominent in the G and Q1 curve.  

The origin of this 2175 \AA\ feature is generally interpreted
as due to the graphite grains component of the interstellar 
dust (Mathis 1990).
In starburst galaxies the absence of this feature was interpreted 
to be the effect of a complex geometry for the dust configuration 
or to a chemical composition different from 
that of our Galaxy (Calzetti et al. 1994). 
For the SMC extinction curve, the absence of the 2175 \AA\ hump was 
explained by a different relative abundances of graphite and 
silicate grains (Pei 1992).

Strictly speaking, the Calzetti et al. (1994) extinction curve
applies to an extended region rather than to a simple line of
sight. Nevertheless we think it is instructive to test it
because the above mentioned peculiarities of its shape (the absence 
of the strong 2175 \AA\ hump and the curvature flatter than 
$\approx \frac{1}{\lambda}$)
seems particularly well suited to describe the afterglow optical-NIR
SEDs.

The optical extinction is typically quantified by the extinction
in the visual band.
Tables \ref{tbl-7} and \ref{tbl-8} give the best fit rest frame 
visual extinction $A_{Vr}$ and the $\chi^2$ for the two 
different spectral index values obtained for $\nu_o<\nu_c$ and 
$\nu_o>\nu_c$ and the four extinction models.  
In all cases the extinction is assumed 
at the redshift of the GRB, or at z=1 when the redshift of the 
GRB is unknown.

In the next Sections individual cases for which interesting results have been obtained from 
our analysis are discussed. 

\subsubsection{GRB 010222 and GRB 000926}

We found that, for these two afterglows, the C extinction 
curve provides the best fit model to the optical SED for 
GRB 000926 and GRB 010222 for $\nu_o<\nu_c$ 
(Figure \ref{f7} and Table \ref{tbl-8}).
The SMC extinction curve provides a fit of quality 
similar to that obtained with the C model for GRB 010222
while it does not provide a good fit to GRB 000926.
Extinction curves with a strong 2175 \AA\ hump provide an unacceptable
fit for these GRB. 
By comparing the G model with respect to the C 
model for the case $\nu_o<\nu_c$, we found a probability  
$P(>\Delta\chi^2)$ of $0.1\%$ for GRB 000926 and 
$P(>\Delta\chi^2)$ of $0.5\%$ for GRB 010222.
We note that these two bursts are those with the widest optical-NIR 
frequency coverage, which strongly helps in constraining the extinction model.

Lee et al. (2001) find acceptable fits for GRB 010222 assuming the 
SMC extinction curve with $A_V\sim0.10$ for $\nu_o>\nu_c$
consistent with our measure (see Table 8).
For the same burst, Masetti et al. (2001) obtain acceptable 
fit over the NIR-optical range using the starburst extinction model,
assuming a continuum model different from that one of Lee et al. (2001), 
for $\nu_o<\nu_c$. Their extinction value is somehow 
higher but still consistent with 
the value that we found for the C model with $\nu_o<\nu_c$.
Bj\"ornsson et al. (2002) derive an optical spectral index of 
$b=1.25\pm0.03$ for $\nu_o<\nu_c$. 
We tested the latter spectral index value and we 
found that it does not affect the result on the C model as the best 
fitting extinction curve for GRB 010222. Galama et al. (2003) found
a rest frame visual extinction value comparable to the Lee et al. (2001) 
result by using the host galaxy extinction curve model by Reichart (2001). 

For GRB 000926 rest frame extinction measured from our analysis 
are consistent with those of Fynbo et al. (2001), Price et al. (2001)
and Harrison et al. (2001) using the Galactic 
or the SMC extinction curves and their derived spectral index. 
However, we point out that a better 
agreement with the data is obtained for this burst with the C 
extinction curve (see Table 8). 

\subsubsection{GRB 971214 and GRB 980519}

The extinction curve that provides the best fit to the SEDs of
these GRB is the SMC model (Figure \ref{f8}).
The statistical improvement with respect to the G model 
for $\nu_o<\nu_c$, is $\Delta\chi^2=13$ for 
GRB 980519 and $\Delta\chi^2=3.6$ for GRB 971214 
(with no NIR points), corresponding to a probability 
$P(>\Delta\chi^2)<0.1\%$ and $P(>\Delta\chi^2)$ of $7\%$ respectively.
We remind that for these two afterglows, the number of photometric 
points is smaller than for GRB 010222 and GRB 000926 and 
therefore the constraint on the extinction model is less tight. 
Moreover, the redshift of GRB 980519 is unknown (we fixed it at z=1). 
We tested also the redshift as a free parameter. 
We found that z$\sim$0.7 improves 
the quality of the fit using the G and Q1 models, 
because the 2175 \AA\ hump moves out 
of the observed frequency range. 
However, the statistical improvement is admittedly low.

For GRB 971214 we tested the hypothesis that the spectral 
break observed between the optical and NIR bands is due to 
extinction (Ramaprakash et al. 1998, see \S 4.3). 
We found that the SMC can fit the data (see Figure \ref{f8}), 
with $A_{Vr}$ consistent with the extinction 
value obtained for the same model fitting only the V, R and 
I points. By comparing the G model with respect to the SMC model for $\nu_o<\nu_c$, we found a probability $P(>\Delta\chi^2)$ of $3\%$.

\subsubsection{GRB 970508}

GRB 970508 is the only case for which the 
extinction curves including a strong 2175 \AA\ hump (G and the Q1) 
provide a good fit to the data (see Table \ref{tbl-7}). 
However, the C and SMC models provide
only slightly higher $\chi^2$.

\subsubsection{GRB 980329}

The redshift of GRB 980329 is unknown, puzzling the interpretation
of the sharp break between the R and I points observed in its SED
 (e.g. Palazzi et al. 1998, Reichart et al. 1999b). 
This spectral jump has been explained in terms of Ly$\alpha$ absorption 
from neutral hydrogen along the line of sight to a high redshift source 
($z>5$ Fruchter et al. 1999b). 
However, Yost et al. (2002) suggested that such a high redshift 
is not compatible with the GRB host galaxy colors.
The sharp break between the R and the I band may be due to the presence
of a strong 2175 \AA\ hump opportunely redshifted.
Indeed, although the $\chi^2$ are still not acceptable, 
we found a minimum $\chi^2$ 
at $z=1.8\pm0.1$ by using the Galactic extinction curve (Figure \ref{f9}).
A similar analysis has been performed by Lamb, Castander \& 
Reichart (1999) on the broad band, from radio to 
X-ray SED and they obtained a qualitatively similar result. 
These values are consistent with the redshift range found 
by Jaunsen et al. (2003) that excluded at $95\%$ confidence level 
$z>4.2$ and $z<1.2$. The latter authors found an host galaxy 
photometric redshift of z$\sim3.6$.

\subsubsection{GRB 970228, GRB 980519, GRB 990123 and GRB 990510}

We found that, for these afterglows,
best fit spectral models are consistent with null extinction
for both the two tested spectral indexes.

The lack of extinction for these afterglows 
will put severe constraints on the estimation of 
the dust to gas ratio in GRB environment (see next Section),
especially the case of GRB 990510, for which a significant 
$N_H$ column density has been detected.

A dedicated spectral analysis of the bright GRB 990123 has been previously 
performed by Galama et al. (1999), Andersen et al. (1999) 
and Holland et al. (2000).
They found spectral index values of $b=1.33\pm0.02$ (from optical to X-ray
spectrum), $b=1.250\pm0.068$ (from a weighted mean over temporal scansion 
of the optical spectrum) and $b=1.31\pm0.10$ respectively, and an extinction
consistent with zero. 
We tested their spectral indexes in our analysis and we obtain 
similar results.

\subsection{$N_H$ versus $A_{Vr}$}

Figures \ref{f10}, \ref{f11} and \ref{f12} compare the best 
fit additional column densities at the redshift of the GRB 
(see previous Section) with best fit $A_{Vr}$ obtained with 
the spectral index $b_1$, corresponding to the case $\nu_o<\nu_c$ 
(see Tables \ref{tbl-7} and \ref{tbl-8}) 
and assuming three different extinction curves.
 
In the case of Galactic extinction curve (Figure \ref{f10}), several GRB lie 
well above the $N_H$ versus $A_{Vr}$ 
relationship expected if a Galactic dust to gas mass ratio is assumed.
The same analysis was performed with spectral index $b_2$, 
corresponding to the case $\nu_o>\nu_c$ and
the obtained best fit extinction values are in this case even lower than
in the previous case (Tables \ref{tbl-7} and \ref{tbl-8}). 
The solid curve in Figure \ref{f10} is the relationship estimated by 
Predehl \& Schmitt (1995) between $N_H$ and $A_{Vr}$: 
$N_H=A_{Vr}\times1.8\times 10^{21}$ cm$^{-2}$. 
We performed a $\chi^2$ test to compare the rest frame $N_H$ 
values (for those afterglows at known redshift) to those 
predicted by the observed extinction $A_{Vr}$ based on this relationship,
and we obtained $\chi^2=12.1$ for 5 d.o.f. 
(probability P($>\chi^2$) of $3.5\%$).

This behavior has been seen in other cosmic sources like AGN as can
be seen from the open squares in Figure \ref{f10} (from a compilation
of Maiolino et al. 2001b and Elvis et al. 1998).  It appears that GRB
are similar to AGN in their $N_H$/A$_{Vr}$ ratio, although AGN show
somewhat higher $N_H$ and A$_{Vr}$.

A good agreement between measured $N_H$ and $A_{Vr}$ is obtained
using the SMC model, since for the SMC, the $N_H/A_{Vr}$ 
ratio is roughly one order of magnitude higher than the Galactic case and
is $N_H/A_{Vr}=1.6\times10^{22}$ cm$^{-2}$ (Weingartner \& Draine 2001).
We performed a $\chi^2$ test to compare the rest frame $N_H$ 
values (for those afterglows at known redshift) to those 
predicted by the observed extinction $A_{Vr}$ 
and we obtained $\chi^2=7.0$ for 5 d.o.f. 
(probability P($>\chi^2$) of $25\%$, Figure \ref{f11}).
We note, however, that in estimating the column densities from the X-ray absorption, a metallicity lower than the Galactic one 
should have been considered if we hypothize a SMC-like environment (Pei 1992).
To reproduce the energy cut-off in the X-ray afterglow spectra
with a metallicity $\sim 1/8$ of the solar one, 
an hydrogen column density increased by a factor 
of $\sim 7$ is required.
We measured the rest frame column densities with a chemical
abundance $\sim1/8$ of the solar one and we compared them to those 
predicted by the observed extinction $A_{Vr}$. We found that
the $\chi^2$ increases by 24.1, with $P(>\Delta\chi^2)<0.001$.

A better result is obtained using the Q1 extinction curve.
For this latter model the 
$N_H$/$A_{Vr}$ ratio is $7.1\times 10^{21}$ cm$^{-2}$ 
(Maiolino et al. 2001a) for a solar 
metallicity, thus providing a self consistent solution.
The $\chi^2$ obtained by comparing the measured $N_H$ to those predicted
by the observed $A_{Vr}$ is $\chi^2=5.4$ for 5 d.o.f. 
(probability P($>\chi^2$) of $35\%$, Figure \ref{f12}). 
We note that the Q1 model does not usually provide good fits to the observed 
optical-NIR SEDs, because of the presence of a strong 2175 \AA\ hump 
(see Figure 6). However, it  seems that this is the model 
that can better explain the measured $N_H$/$A_{Vr}$ values.  
This is due to the flat shape of this extinction curve 
that gives a more gray extinction than the Galactic or SMC cases
and to the high $N_H/A_{Vr}$ ratio.  

The C extinction curve has a similar shape to Q1 (see Figure \ref{f6}) and 
no strong 2175 \AA\ hump. Unfortunately, the $N_H/A_{Vr}$ ratio has 
not been estimated for the starburst galaxies, because of the complexity 
of the geometry of the dust and stars distribution  
inside these galaxies (Calzetti 2001). Therefore, we 
cannot test the data against the prediction of this model.

\subsubsection{Test of the fireball model}

As discussed in section 5.2 the continuum spectral index adopted
in the optical-NIR photometry fits was derived from the best fit X-ray
spectral index, assuming the fireball model, and assuming that the
X-ray frequencies are above the cooling frequency.  We have performed
a selfconsistency check of these assumptions by plotting the afterglow
NIR to X-ray SED, after correction for dust extinction and gas
absorption, along with the best fit X-ray power law model and the
assumed optical power law model. The normalization of the X-ray
spectra has been scaled to the same time of the optical-NIR
observations, using the published decay indices.  We note that in most
cases (but GRB 970228 and GRB 970508) 
the X-ray observations encompass the times at which the
optical-NIR observations were performed, see Table 3 and 4, thus
minimizing any systematic uncertainty in the scaling of the X-ray
spectra.  Figure \ref{f13} shows three examples: GRB 990123, GRB
990510 and GRB 010222, which are among the brightest of our sample.
In these three cases the optical-NIR photometry has been corrected for
rest frame extinction using the best fit value relative to the Q1
model, or its upper limit when it is consistent with zero. 
Similar results are obtained using the C model. Note as using these
extinction laws the optical and X-ray spectra of GRB 010222 are 
consistent with the fireball model. Adopting the G and the SMC 
extinction laws, or no extinction, we find again the
result already reported by in't Zand et al. (2001).
Figure \ref{f13} show that the NIR to X-ray SEDs are well consistent with the
fireball model, and that the cooling frequency is constrained within the
UV and the X-ray band. Qualitatively similar results are found also for the
other afterglows. We remark, however, that in latter cases this is not
a truly stringent test, given the lower statistics in the X-ray spectra 
and the consequent larger uncertainties on the X-ray spectral index, 
(and therefore also in the allowed range of values for the optical index).

\section {Discussion}

\subsection{X-ray analysis}

We analyzed the X-ray absorption properties of a sample of
13 bright GRB afterglows observed by BeppoSAX in the energy range 
$0.1(0.4)-10.0$ keV.  We found
that absorption in addition to the Galactic one 
along the line of sight is highly statistically significant in two cases: 
GRB 990123 and GRB 010222 (probability $>99.9\%$). 
These are also the two GRB with the best statistics.
However, fitting the spectra of two other GRB afterglows 
with lower but still reasonably good statistics (GRB 971214 and GRB 970228), 
including an additional absorber does not produce a significant improvement 
in $\chi^2$ (probability of $93.3\%$ and $90.8\%$ respectively).
In  three other cases (GRB 980703, GRB 990510 and GRB 001109) 
the presence of an additional absorber is marginally
significant (probability $\sim97\%$).
X-ray absorption may be common,
although the quality of present data does not allow us to reach a firm
conclusion. In all cases, the absorbing columns at the GRB 
redshift are in the range $N_H=10^{21}$ - a few $\times 10^{22}$ cm$^{-2}$, 
values similar to the column usually observed toward the disk 
and the bulge of the Galaxy, suggesting that GRB
afterglows are probing similar environments in the GRB host galaxies.
This range is also in agreement with the estimation of the $N_H$
expected if GRB were embedded in giant molecular clouds similar to 
those observed in our galaxy (Reichart \& Price 2002).

De Pasquale et al. (2003) performed an analysis on the 
BeppoSAX X-ray afterglows, aimed at studying the nature of Dark GRB 
(for which no or very faint optical counterpart has been discovered). 
Their conclusions from column density measures are 
generally consistent with those given in this paper.

The rest frame column density estimated
for 9 X-ray afterglows observed by XMM (GRB 011211, GRB 020322 and GRB 030227) 
and Chandra (GRB 991216, GRB 000210, GRB 000926, GRB 020405, GRB 020813 
and GRB 021004) are consistent within the range found for the
13 BeppoSAX GRB (see Figure \ref{f3}). 

The spectrum of GRB 010222 suggests the presence of complex absorber
or an additive emission component (Figure \ref{f5}).
Two possible explanations invoke the recovery of the spectrum 
below the energies of strong ionized absorption edge lines as OVII, OVIII 
or the spilling of the emission spectrum through a leaky absorber.
We tested both models obtaining a significant improvement of the
$\chi^2$ in both cases (Figure \ref{f5}). We found an 
upper limit to the equivalent width of a line
at 6.4/(1+z) keV of 150(1+z) eV (90\% confidence level),
that is, one order of magnitude smaller than the line equivalent width 
measured for GRB 000214 (Antonelli et al. 2000). 

\subsection{Optical-NIR SEDs fitting}

We evaluated the rest frame visual extinction
of the nine optical afterglows in our sample. 
In addition we analyzed also the optical afterglow of GRB 000926 
(see \S 5.2.1).

We fitted the optical-NIR afterglows photometry
with a power law model modified at short wavelengths 
by rest frame extinction (see equation (1) in \S 5.2).  
We tested four different extinction curves: {\it i)} Galactic 
 (``G''); {\it ii)} Small Magellanic Clouds 
(``SMC''); {\it iii)} the extinction curve computed for a 
sample of local starburst galaxies (``C''); 
{\it iv)} an extinction curve computed from simulations by 
Maiolino et al. (2001a) based on a dust distributions
skewed toward large grains (``Q1'').

In two cases, GRB 010222 and GRB 000926, we found that the 
extinction curve that better fit the data is the C model. 
These afterglows have the 
widest optical-NIR frequency coverage in our sample. 
This strongly help in constraining the extinction model.
In particular, those extinction curves with a strong 
2175 \AA\ hump (G and Q1) are not suitable to model these two afterglows.
In fact, despite Q1 has a slope very similar to that of the C model 
(see Figure \ref{f6}), it
fails to fit their optical SEDs due to the presence 
of a strong 2175 \AA\ hump that, on the other hand is completely
absent in the C curve.

We note that assuming the C or the Q1 model, the NIR to X-ray SED of
GRB 010222 is fully consistent with the fireball model, 
without the need of a strong contribution from inverse Compton scattering
in the X-ray band (see eg. in't Zand et al. 2001).

The SMC model provides good fit to the GRB 980519 
and GRB 971214 data (by using for the latter both optical 
and NIR data, see \S 5.2.2). 
However, the number of photometric points is smaller than for 
GRB 010222 and GRB 000926 and 
therefore the constraint on the extinction model is less tight.
Moreover, the redshift of GRB 980519 is unknown and for GRB 971214 
the interpretation of the spectral break observed between 
the optical and NIR bands is not univoque  
(see \S 4.3). Excluding the NIR points, both the SMC and the
C extinction curves provide comparable good fit to GRB 971214.

The small curvatures of the GRB 990510 and GRB 990123 SEDs require
very low extinction. No strong constraints on the extinction model 
can be obtained also in the cases of GRB 970228 and 
GRB 980329 because of the few photometric points.

\subsubsection{The 2175 \AA\ hump}

The extinction models tested in our analysis 
that have a strong 2175 \AA\ hump are indicated as G and Q1 
(see \S 5.2 and Figure \ref{f6}).

We found that in all the analyzed optical-NIR SEDs, except GRB 970508, 
the G and Q1 model give bad $\chi^2$ (see Table \ref{tbl-8}).

This 2175 \AA\ hump does not appear to be a common feature 
in extra galactic sources (Calzetti et al. 1994) and there 
is not yet a clear and univoque interpretation of its absence.
In the case of GRBs, the absence of the 2175 \AA\ 
can be explained by the results obtained by Perna et al. (2002b). 
They find that dust destruction of the small graphite 
grains due to an intense X-ray/UV source, as could be for GRB, 
would produce a significantly suppression of the 2175 \AA\ hump. 

Two bursts in our sample are at unknown redshift: GRB 980519 and GRB 980329.
We found that for GRB 980519, a lower redshift (z$\sim$0.7) 
improves, although not very significantly, the quality of the fit 
using the G and Q1 models, 
because the 2175 \AA\ hump moves out of the observed frequency range. 
For GRB 980329, the jump between the R and I photometric 
points is marginally explained by the 2175 \AA\ hump redshifted 
to the optical range (z$\sim$1.8, see Figure \ref{f9}).

\subsection{$N_H$ versus $A_{Vr}$ relationship}

As discussed in the previous sections, we have evaluated rest
frame optical extinction $A_{Vr}$ from the optical to near infrared
photometry and rest frame equivalent hydrogen column densities $N_H$
from the X-ray data.

We found that several GRB lie well above the
$N_H$ versus $A_{Vr}$ relationship expected for the Galactic
extinction curve (Figure \ref{f10}).  In this regard our results are
qualitatively similar to those of Galama \& Wijers (2001), although the
best fit $N_H$ values are somewhat smaller than those
reported by the latter authors, and visual extinction are evaluated here
using detailed extinction curves (while Galama \& Wijers (2001) 
assume a linear extinction curve $A(\lambda)\propto
\frac{1}{\lambda}$ and computed rest frame $N_H$ by using a simple 
power law scaling of the $N_H$ measured in the observer frame).

Scenarios that can reconcile the observed X-ray absorption with
the lack of strong optical extinction, involve the presence 
of a dust to gas ratio much lower than the Galactic one (see
Figure \ref{f6}), and/or of a dust
grain size distribution skewed toward large grains. 
We discuss these two possibilities in turn.

A dust to gas ratio much lower than the Galactic one has been
tested assuming the SMC extinction curve. In fact, for the SMC ISM it
has been inferred a dust content $\sim1/10$ of the Galactic one (Pei
1992).  A good agreement between the best fit $N_H$ and $A_{Vr}$ is
obtained using this extinction curve (Figure \ref{f11}).  We note, however,
that the best fit $N_H$ values have been obtained by assuming Galactic
metal abundances, while the metallicity of the SMC ISM is $\sim 1/8$
the Galactic one (Pei 1992).
If such low metallicity were assumed, best fit $N_H$ values became 
$\approx 7$ times higher than those quoted in Tables 5
and 6 as plotted in Figure \ref{f11} (dashed line),
worsening the agreement with the best fit
optical extinction.  We can therefore conclude that also for an
SMC-like ISM the observed extinction is significantly lower than
predicted from the X-ray column density.

Large grains give gray opacity from optical to the infrared 
(e.g. Kim, Martin \& Hendry 1994, Weingartner \& Draine 2000, 
Maiolino et al. 2001a), that could explain the small curvature
observed in the optical afterglow SEDs. 
A low $A_V/N_H$ supports the scenario were large grains
are formed by coagulation mechanisms (e.g. Kim \& Martin 1996, 
Maiolino et al. 2001a).
Since the grain coagulation rate increases with
density as $n^{1/2}$ (Draine 1985),
the depletion of small grains by coagulation mechanisms is 
expected to be particularly efficient in high density environments, 
as the giant molecular clouds which could harbor GRB events.

High density gas and dust coagulation is not the only viable 
mechanism to produce a dust grain size distribution skewed 
toward large grains.  
If the dust to metal ratio is constant, as in the Galaxy and 
the two Magellanic Clouds ISM (Pei 1992), the lack of strong 
extinction (dust), along with a consistent amount of X-ray 
absorption (metal), suggest a
scenario were dust is destroyed by the intense GRB X-ray/UV flux. 
Dust grains may be heated and evaporated by the intense X-ray and UV
radiation field up to $\sim20pc$ (Waxmann \& Draine 2000, Draine \&
Hao 2002, Fruchter, Krolik \& Rhoads 2001).  Perna et al. (2002b) and
Perna \& Lazzati (2002c) show that the result of the exposure of dust
to the intense GRB radiation field can be a grain size distribution
flatter that the original one.  The main reason is that dust
destruction is more efficient on small grains.  Perna et al. (2002b)
computed the extinction curve which is obtained if standard Galactic
dust is exposed to a GRB, and found that the dust extinction curve can
be very flat, at least for burst lasting more than a few tens of
seconds.  

We tested a dust model skewed toward large grains using the
model Q1 adapted from Maiolino et al. (2001a).  The X-ray $N_H$ agree
very well with the hydrogen column density expected from the best fit
$A_{Vr}$ and the visual extinction-reddening ratio appropriate for
this model (see Figure \ref{f12}). We note that this model does not imply a
metallicity lower than solar, thus providing a self consistent
solution, unless the SMC-like ISM. 

However, the Q1 curve gives generally worse $\chi^2$ (see Table 7 and
8) than the SMC and C models. This is mainly due to the presence of
a strong 2175 \AA\ hump in the Q1 model.  The C extinction curve has
a slope similar to that of the Q1 model (Figure \ref{f6}), but no 2175 \AA\
hump, thus providing better fits.  This difference does not affect the
rest frame extinction estimates, whose values from both models are
indeed comparable (see Table 7 and 8) and suggests that an 
extinction model with a ``flat'' slope and no 
2175 \AA\ hump would produce both good fits to most observed 
SEDs and a nice agreement between $N_H$ and $A_{Vr}$.

In the near future we will have the chance to confirm this intriguing
scenario. {\it Swift} will enable us to obtain high quality X-ray
spectra of the afterglow as a function of the time. The {\it Swift}
optical monitor will allow us to obtain precise and reliable
optical-UV photometry.  Robotic telescopes like REM, triggered by {\it
Swift}, will be able to obtain simultaneous near infrared
photometry. Therefore it will be possible to obtain precise reddening
measures, constraining the correct extinction law, 
and to follow the evolution of the reddening as a function of
the time, that is the smoking gun of a possible 
GRB UV/X-ray flux influence on their environment.

\acknowledgments

This research has been partially supported by COFIN 2001 grants, by
CNAA 1999 and 2000 grants and by University of Rome ``La Sapienza''
support.

We would like to thank Roberto Maiolino for his precious support to
this work, Rosalba Perna, Sandra Savaglio, Daniela Calzetti, 
Nicola Masetti for useful discussions and an anonymous referee for 
corrections that improved the presentation. BeppoSAX was a program of the
Italian Space Agency (ASI) with participation of the Dutch space
agency (NIVR). We would like to thank all the members of the BeppoSAX
team for the support in performing observations with the satellite.
We also thank an anonymous referee for comments that helped in improving 
this presentation.

\appendix 
\section{Appendix}

{\bf GRB 970228}:
A large number of papers report photometric observations of the
optical afterglow of GRB 970228 (e.g. Guarnieri et al. 1997, Sahu et
al. 1997, Galama et al. 1997, 
Reichart 1997, Pedichini et al. 1997, Garcia et al. 1998, Castander \& Lamb 
1999, Fruchter et al. 1999a, Reichart 1999a, Galama et al. 2000).  The latter paper presents a complete
and exhaustive summary of the available optical observations, and
therefore we refer to this paper for this afterglow.
The Galactic extinction in the direction of GRB 970228 is
$A_V=0.78\pm0.12$, from the SFD maps. Castander \& Lamb (1999)
estimate a Galactic extinction of $A_V=1.09^{+0.10}_{-0.20}$ combining 
the results of four different methods for measuring the extinction
(also see Fruchter et al. 1999a, and Gonzalez,
Fruchter \& Dirsch 1999).  We used both these values and we 
verified that their difference
does not affect significantly the estimation of the rest frame
extinction.

{\bf GRB 970508}:
Galama et al. (1998a) collect observations over the entire
optical band, from U to I, for this afterglow, and therefore we referred
to this paper in this case.
Different authors have reported different estimates of
the Galactic extinction in the direction of GRB 970508.  
Galama et al. (1998a) report $A_V<0.01$ (Laureijs
1989), while Djorgovski et al. (1997) report $A_V=0.08$ (Laureijs,
Helou \& Clark 1994), both using the $100\micron$ cirrus flux maps. 
Reichart (1998) reports $A_V=0.09$, 
using the Rowan-Robinson et al. (1991) maps (also see Sokolov et al. 
1998a, Galama et al. 1998b, Bloom,
Djorgovski \& Kulkarni, 1998a, Castro-Tirado \& Gorosabel 1999).
We adopt here the value found by the two latter authors.

\begin{figure}
\plotone{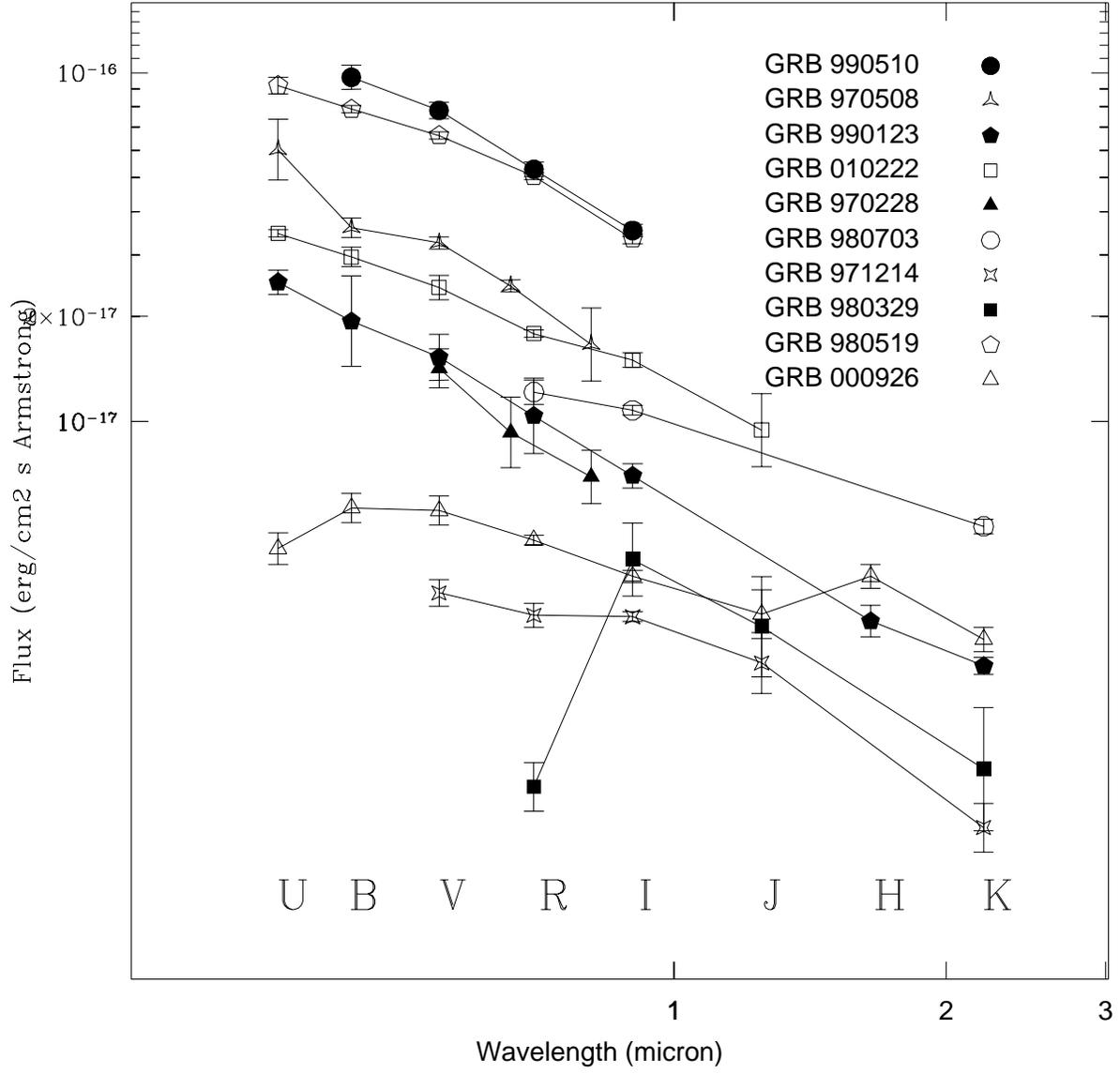}
\caption{Spectral Energy Distribution for 9 of the 13 X-ray afterglows which have optical counterpart, plus GRB 000926 (see Appendix and Table 4) \label{f1}}
\end{figure}

\begin{figure}
\plotone{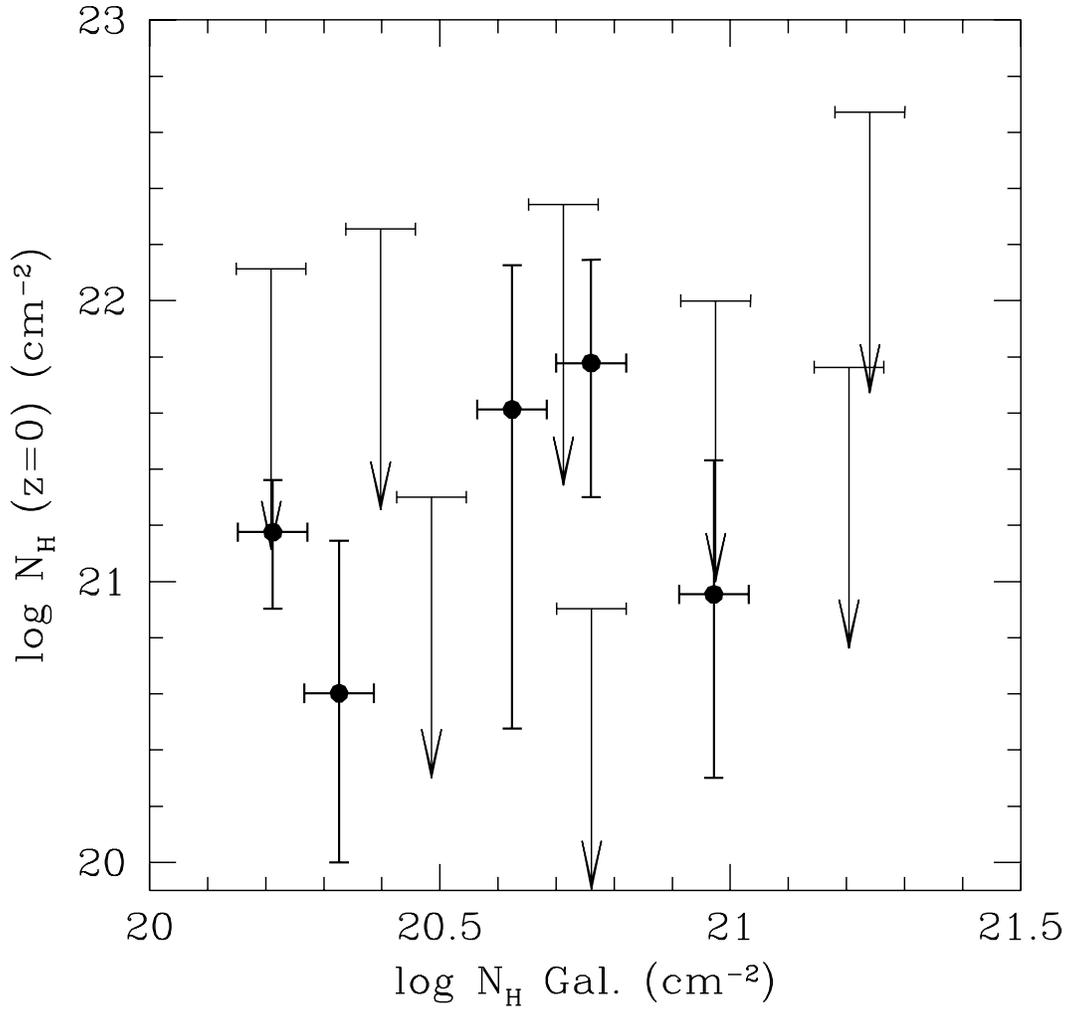}
\caption{Best fit additional column density at z=0 against
the Galactic column. Errors and upper limits (arrows) are at $90\%$ confidence 
level for two parameters of interest ($\Delta\chi^2=4.61$). 
No trend is evident, supporting the robustness of our analysis (see \S 5.1). \label{f2}}
\end{figure}

\begin{figure}
\plotone{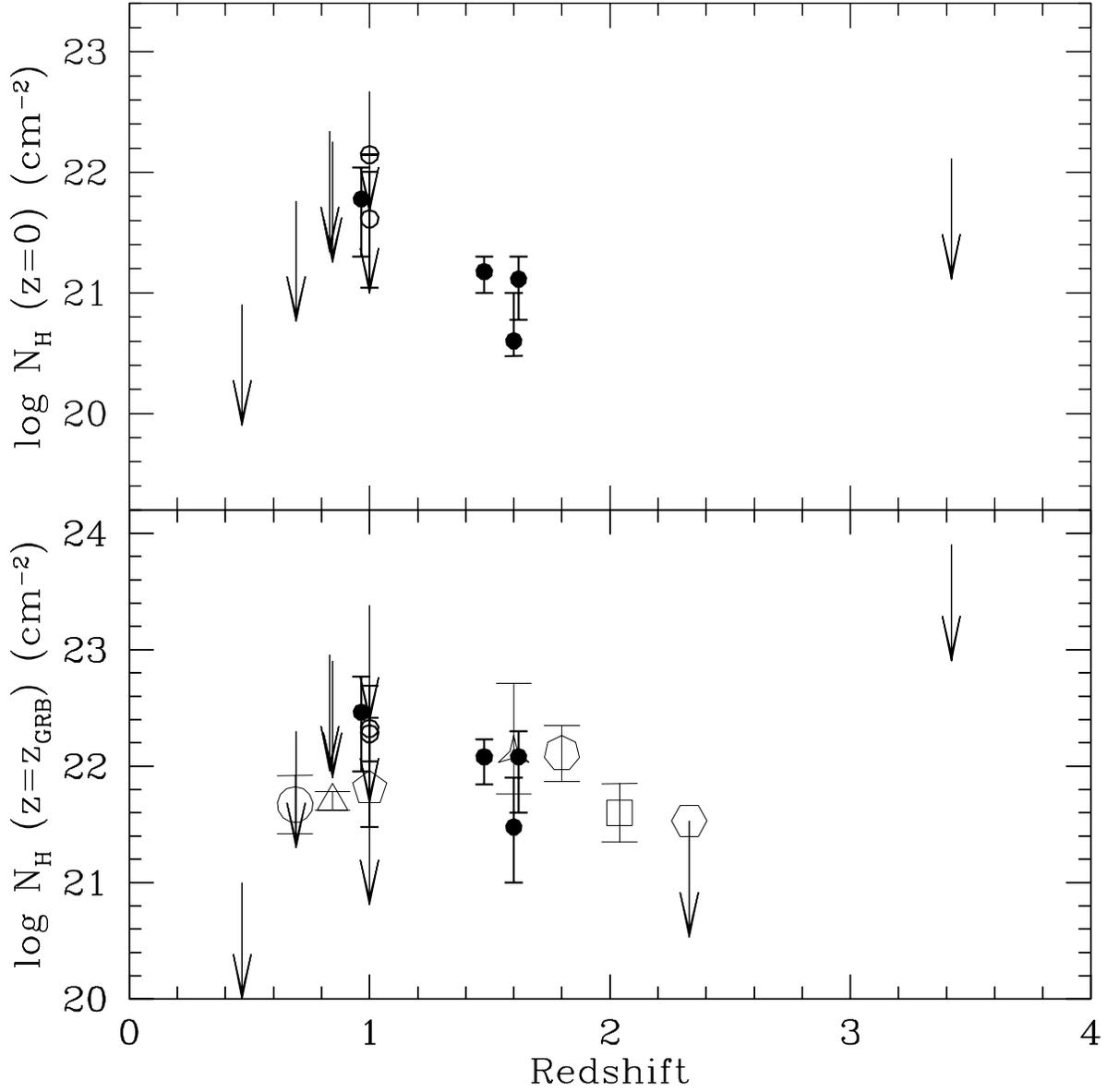}
\caption{
{\it Upper panel}: Additional column density at z=0 versus redshift. {\it Lower panel}: Additional column density at z=z$_{GRB}$ (filled small circles) or
z=1 for GRB with unknown redshift (open small circles) versus redshift. 
Errors and upper limits (arrows) are at $90\%$ confidence level for two parameters of interest ($\Delta\chi^2=4.61$). 
In this panel are also shown the rest frame column densities estimated
for GRB 000926 at z=2.04 (open square) and GRB 000210 at z=0.846 (open 
triangle)  both from Piro et al. (2001,2002b), GRB 991216 at z=1.0 
(open pentagon upper limit) from Ballantyne et al. (2002),
GRB 020322 at z=1.8 (open eptagon) from Watson et al. (2002), 
GRB 020405 at z=0.690 (open circle) from Mirabal et al. (2002b),
GRB 021004 at z=2.33 (open esagon upper limit) from Holland et al. (2002) 
and GRB 030227 at z$\sim1.6$ (open starred triangle) from Mereghetti et al. (2003) and from Watson et al. (2003). \label{f3}} 
\end{figure}

\begin{figure}
\plottwo{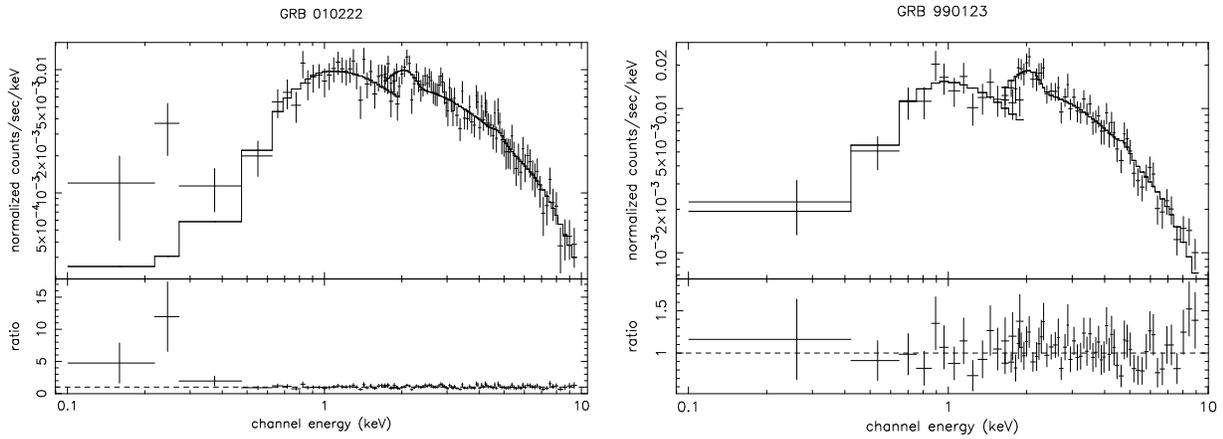}{f4b.eps}
\caption{{\it Left:} LECS and MECS X-ray spectra of GRB 010222 {\it Right:} LECS and MECS X-ray spectra of GRB 990123. Both spectra are fitted by a simple power law and two neutral absorption components, one fixed to the Galactic value and the other free to vary at the redshift of the host galaxy. The residuals at low energies for GRB 010222 suggest the presence of complex absorption or an additive emission component.\label{f4}}
\end{figure}

\begin{figure}
\plotone{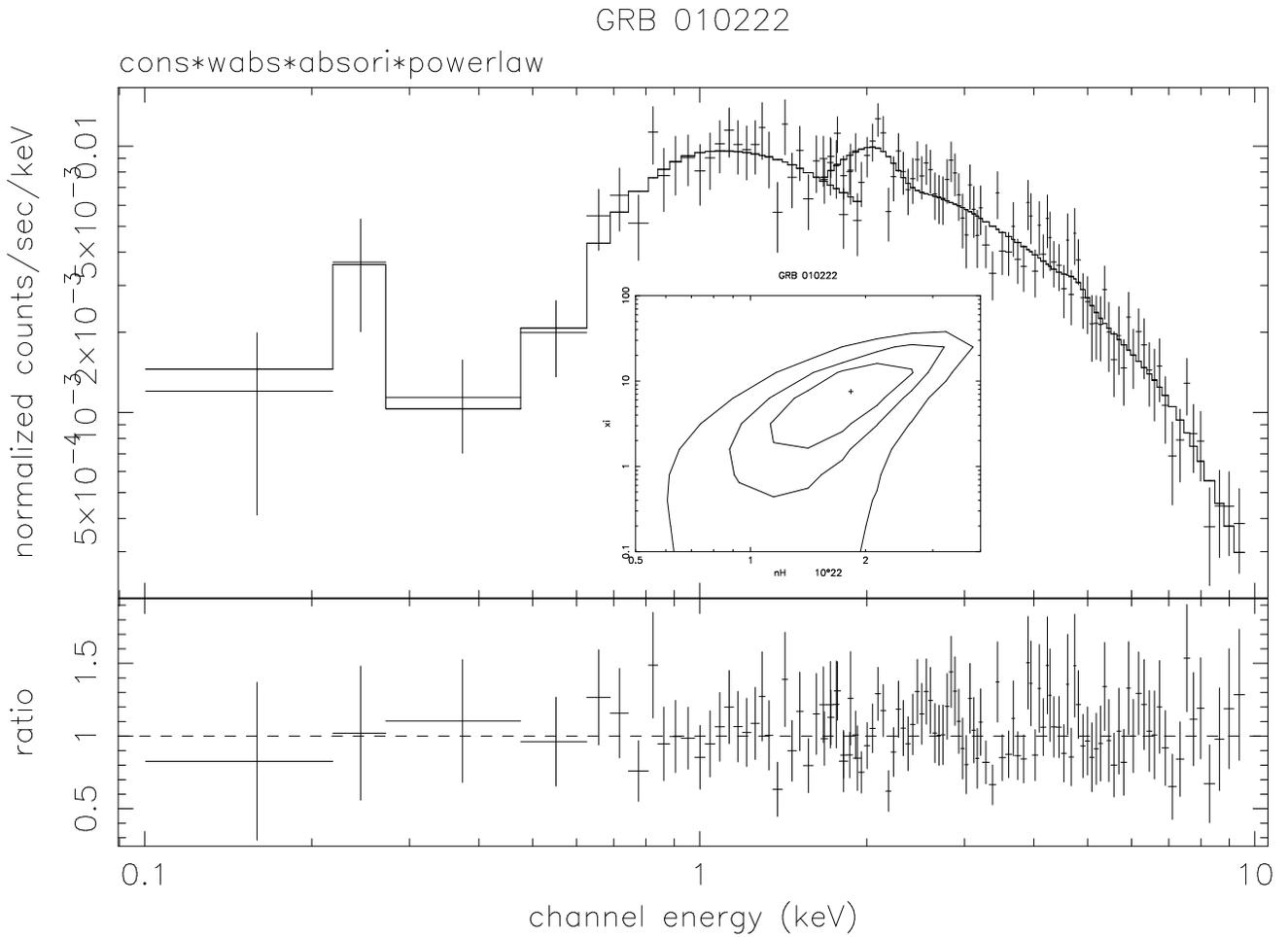}
\caption{LECS and MECS X-ray spectra of GRB 010222 fitted with an ionized absorber (``absori'' in XSPEC, see \S 5.1.1). The insert shows the $\chi^2$ contours of the ionization parameter $\xi$ and the rest frame column density $N_H$ at 65\%, 90\% and 99\% confidence levels.\label{f5}}
\end{figure}

\begin{figure}
\plotone{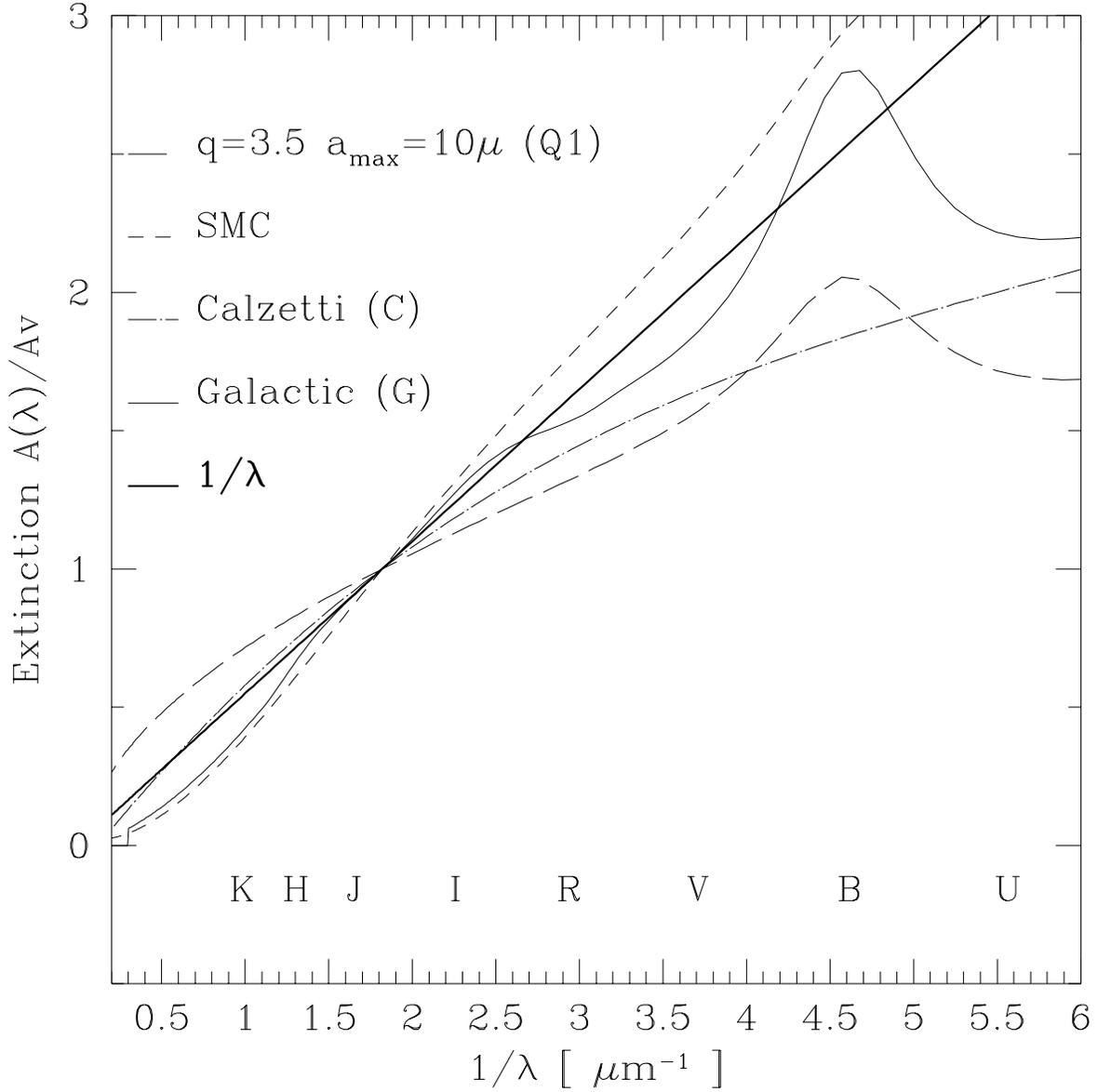}
\caption{Extinction curves adopted in the analysis of the optical-NIR photometry against $1/\lambda$ (see \S 5.2). We indicate in the plot the rest frame wavelengths actually observed by a standard photometric system assuming z=1. 
\label{f6}}
\end{figure}

\begin{figure}
\plottwo{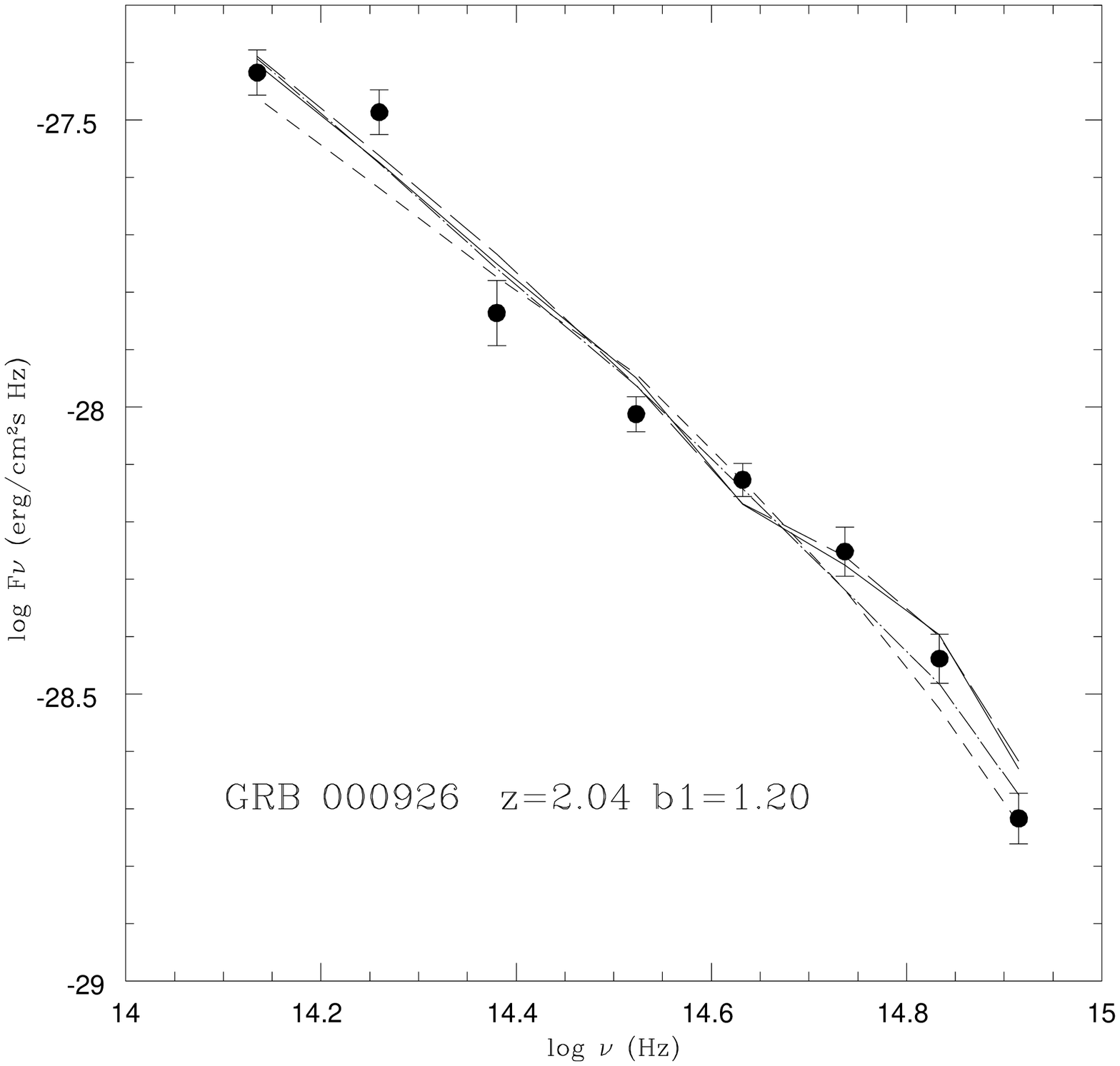}{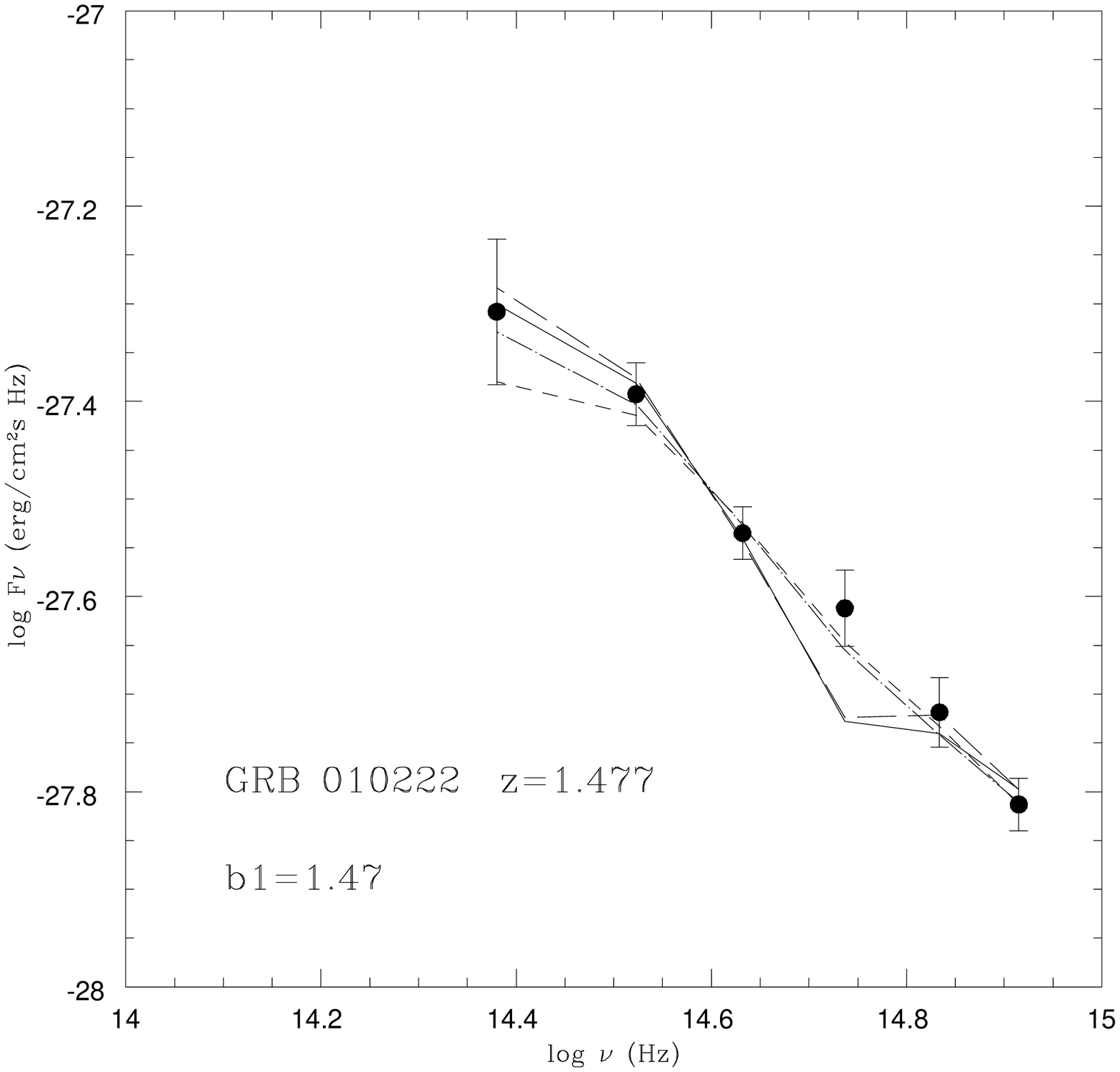}
\caption{{\it Left:} The SED of GRB 000926 fitted with the four tested
extinction models (line code is the same as in Figure \ref{f6}).
{\it Right:} The SED of GRB 010222 with the four tested
extinction models. The 2175\AA\ hump, typical of the extinction 
curves G and Q1, does not fit to GRB 010222. 
The C model (dot-dashed line) provides the minimum 
$\chi^2$ among all the fitted 
models for both these bursts for $\nu_o<\nu_c$ (see \S 5.2.1 and Table 8). 
\label{f7} }
\end{figure}

\begin{figure}
\plottwo{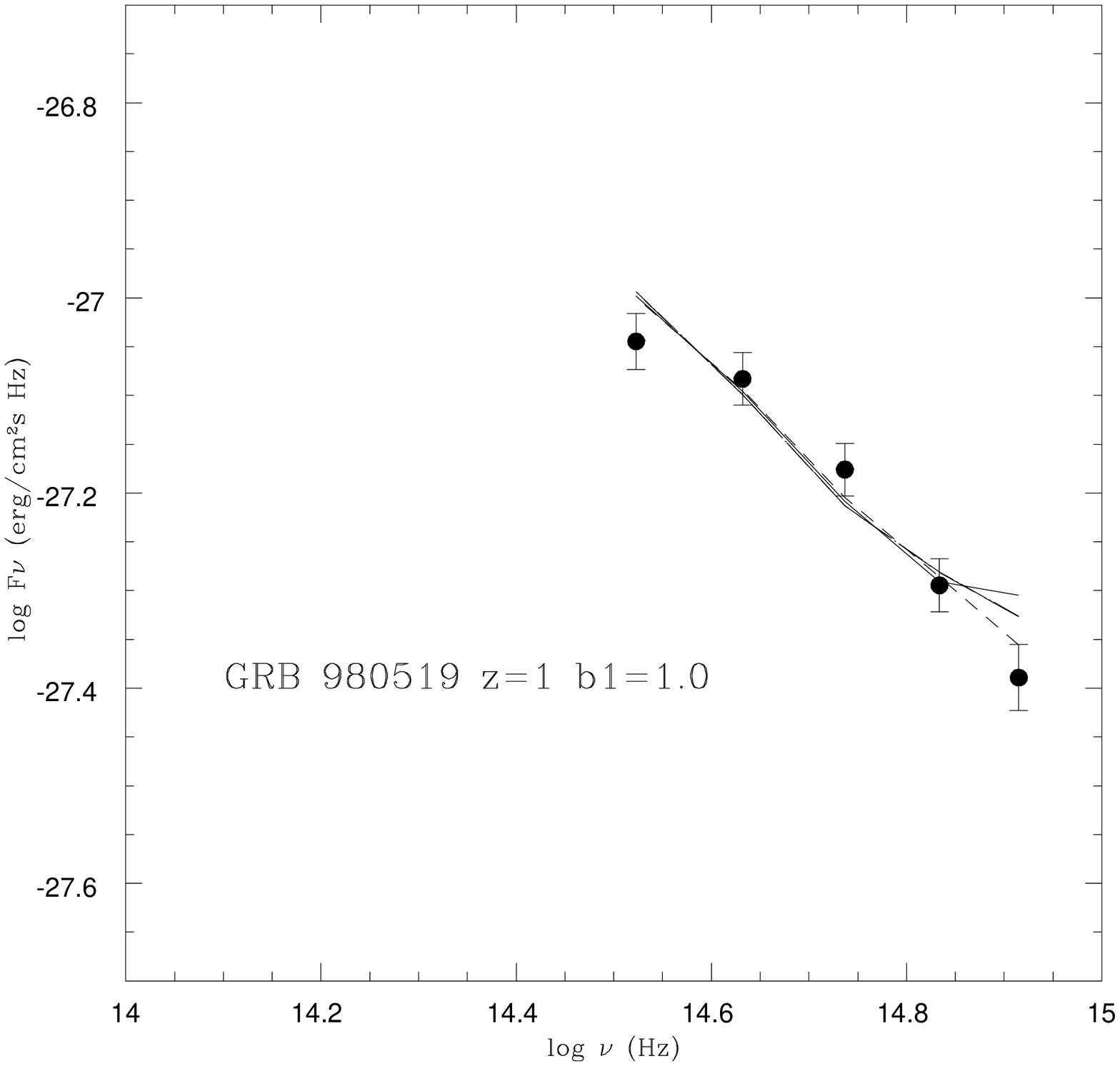}{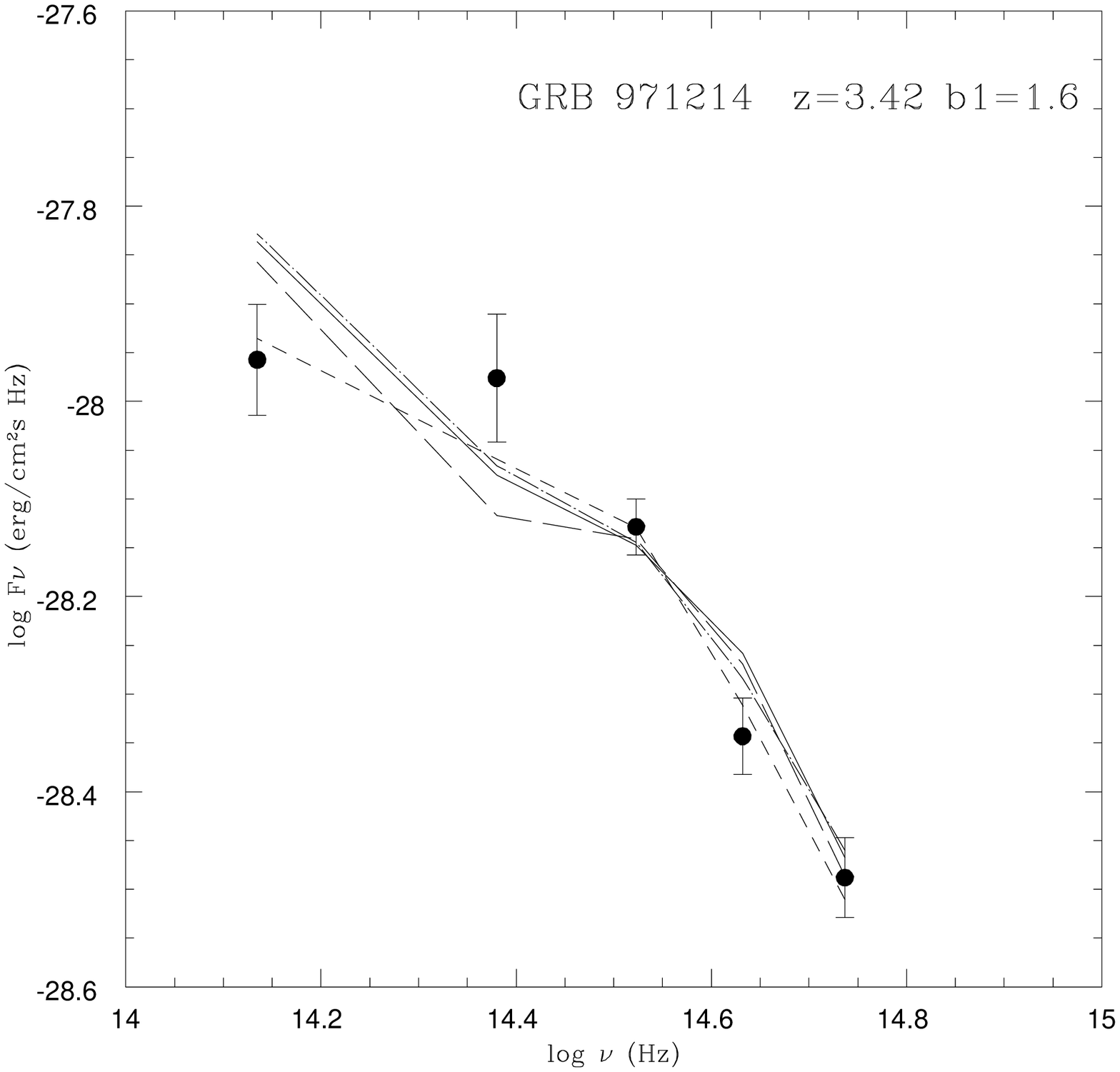}
\caption{{\it Left}: The SED of GRB 980519 with the best fit models 
(line code is the same as in Figure \ref{f6}). The SMC  
extinction curve (short dashed line) well fit to the data.
In particular the extinction curves with the strong 2175 \AA\ hump G and Q1,
 are not suitable to describe its SED at z=1. 
For lower redshift, the 2175 \AA\ is shifted outside the 
photometric range of this burst. We found a minimum $\chi^2$ for z=0.7 by
using both the G and the Q1 extinction curves.
{\it Right}: The SED of GRB 971214 with the best fit models 
(line code is the same as in Figure \ref{f6}).
The SMC extinction curve (short dashed line) provides the 
minimum $\chi^2$ (see Tables \ref{tbl-7} and \ref{tbl-8}). 
\label{f8} }
\end{figure}

\begin{figure}
\plottwo{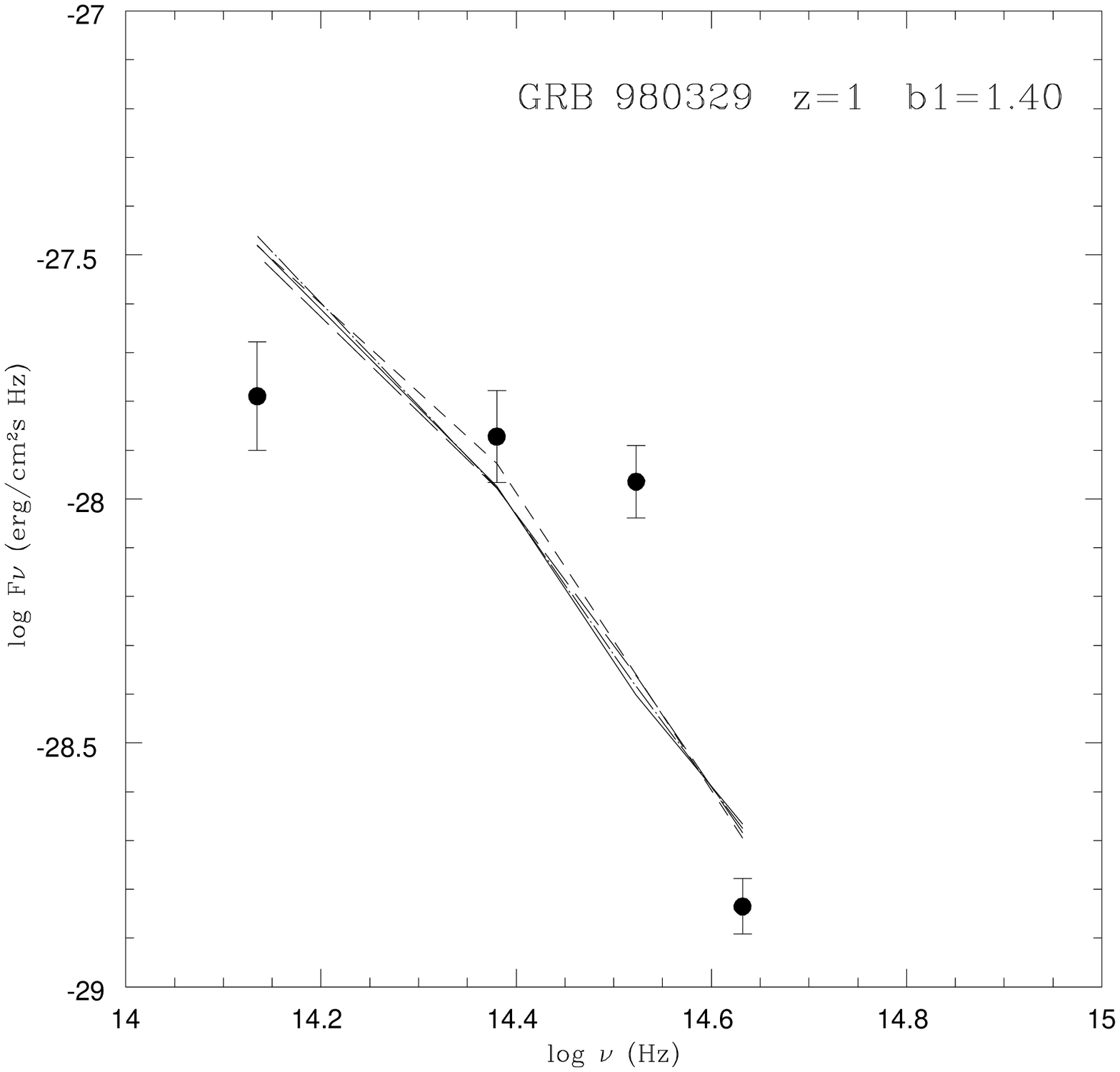}{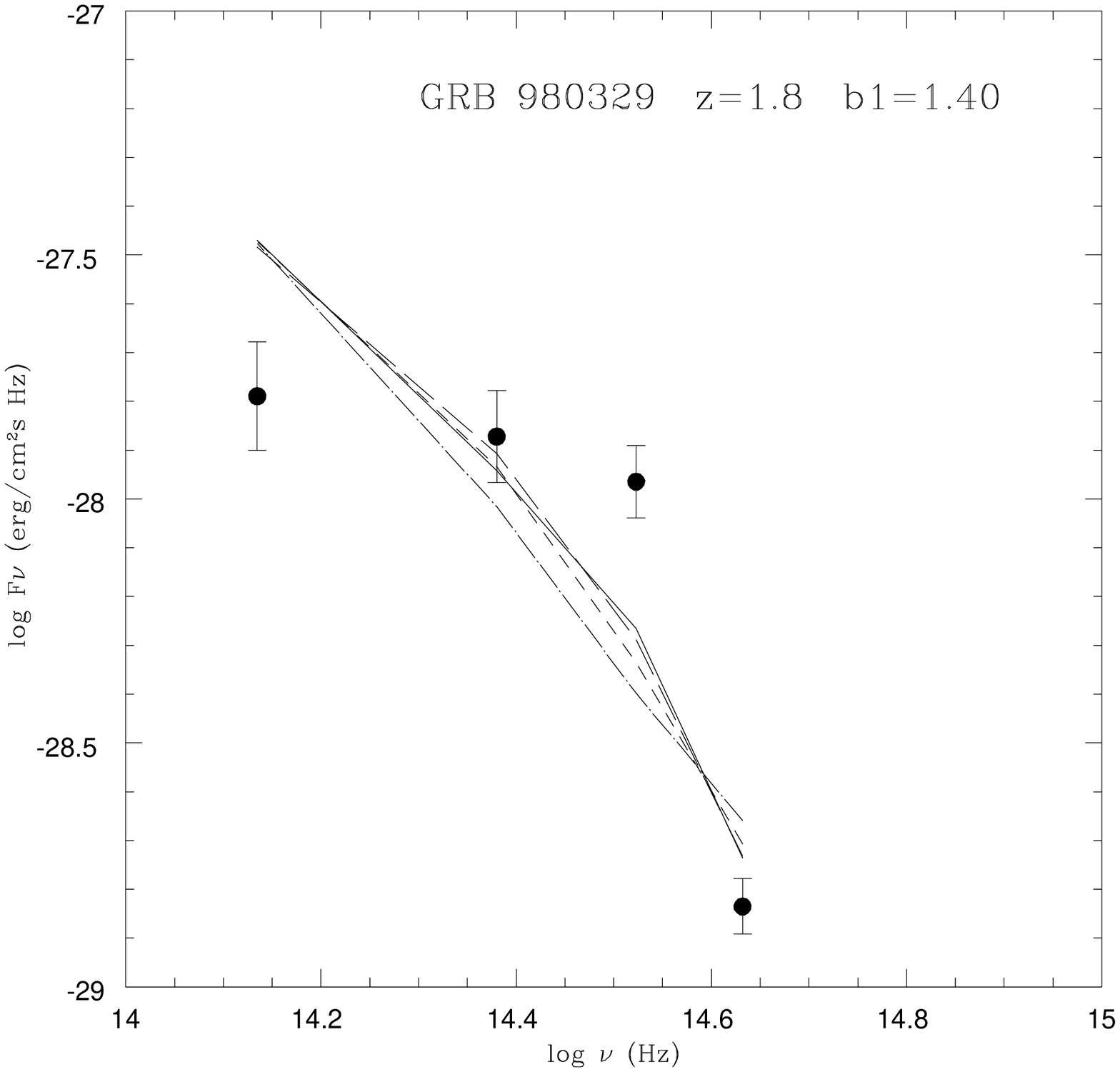}
\caption{{\it Right.} The SED of GRB 980329 with the best fit models 
(line code is the same as in Figure \ref{f6}) fixing the redshift z=1. 
{\it Left.} The SED of GRB 980329 with the best fit models obtained 
leaving the redshift free to vary. A minimum $\chi^2$ is obtained at z=1.8 
for all models. 
The jump between the R and I photometric points is marginally explained by the 2175 \AA\ hump redshifted to the optical range (see \S 5.2.4).
The same result is obtained for both the spectral index $b_1$ and $b_2$ 
(see Tables 7) \label{f9}}
\end{figure}

\begin{figure}
\plotone{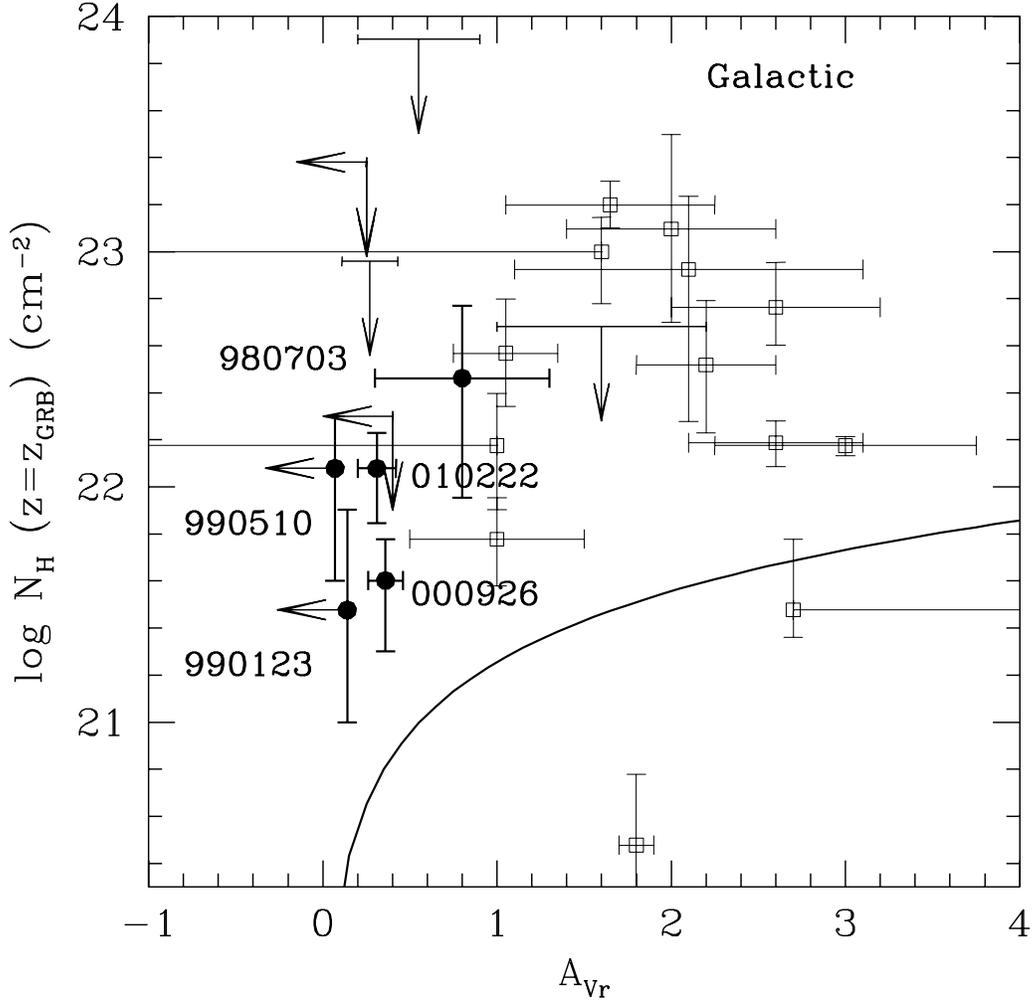}
\caption{
Best fit additional column densities (at the redshift of the GRB or at
z=1 for the GRB with unknown redshifts) against best fit rest frame
visual extinction $A_{Vr}$ obtained assuming the Galactic (G) 
extinction curve (Cardelli et al. 1989). 
The $A_{Vr}$ are estimated using the $b_1$ ($\nu_o<\nu_c$) optical 
afterglow power law indices (see Tables 7 and 8). Error bars are 68\% confidence 
intervals. Upper limits are 90\% confidence intervals.  
Small open squares are a sample of quasars from Maiolino et al. (2001b) and (Elvis et al. 1998).  The solid line curve shows the expected values assuming a Galactic dust to gas mass ratio and the Galactic extinction model 
(Predehl \& Smith 1995). \label{f10}}
\end{figure}

\begin{figure}
\plotone{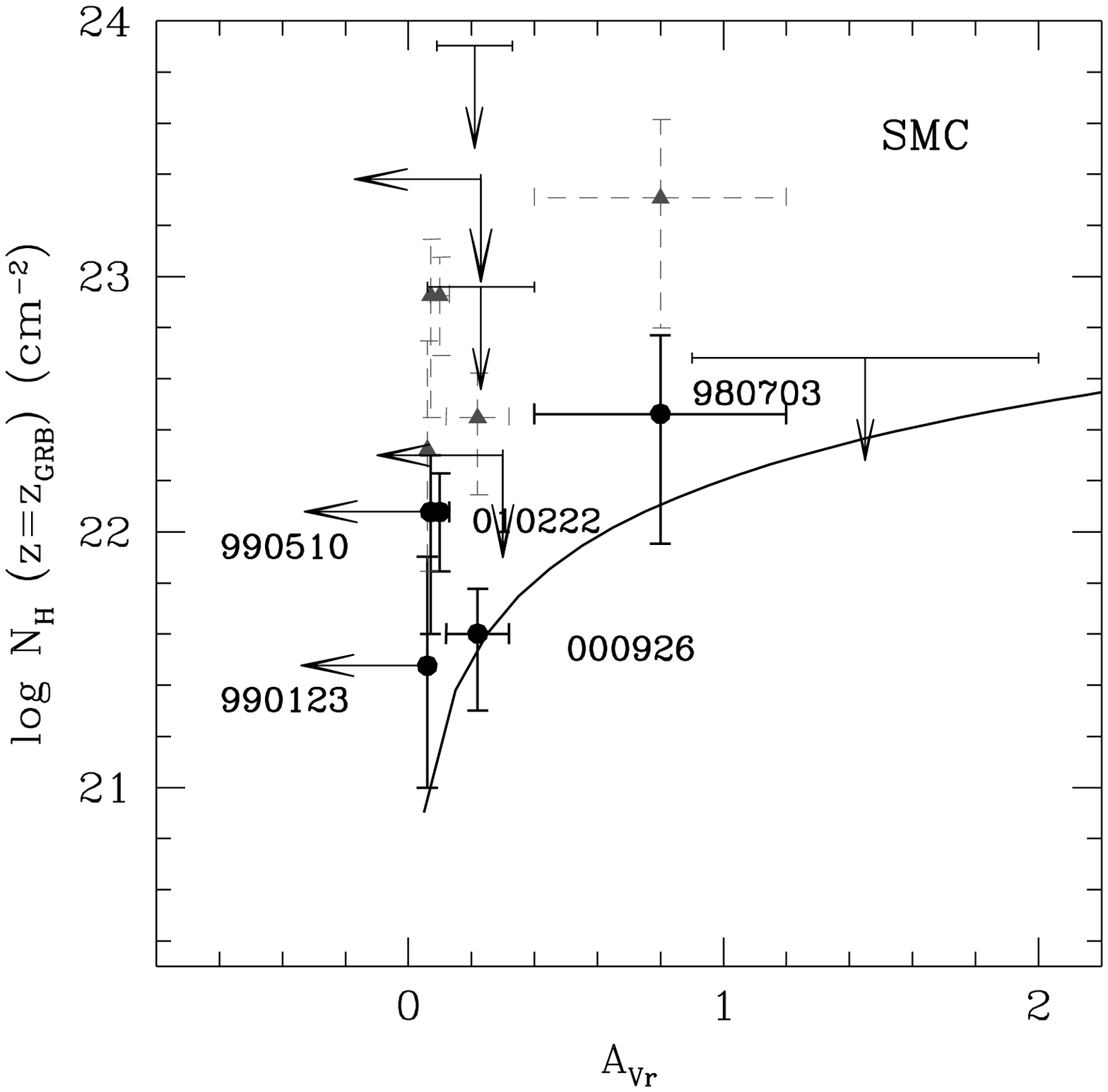}
\caption{
Best fit additional column densities against best fit visual extinction $A_{Vr}$ (both at the redshift of the GRB or at z=1 for the GRB with unknown redshifts) obtained using the SMC extinction curve (Pei 1992). 
The $A_{Vr}$ are estimated using the $b_1$ ($\nu_o<\nu_c$) optical 
afterglow power law indices (see Tables 7 and 8). Error bars are 68\% confidence intervals and upper limits are 90\% confidence intervals. 
The black dots with solid line error bars are the $N_H$ measured 
assuming a solar metal abundance. The triangles with dashed line 
error bars are the $N_H$ measured assuming a SMC environment 
which has a metal abundance $\sim 8$ times lower than the solar one.
The solid line curve is the expected relationship between $N_H$ and 
$A_{Vr}$ for the SMC (Weingartner \& Draine 2001).  \label{f11}}.
\end{figure}

\begin{figure}
\plotone{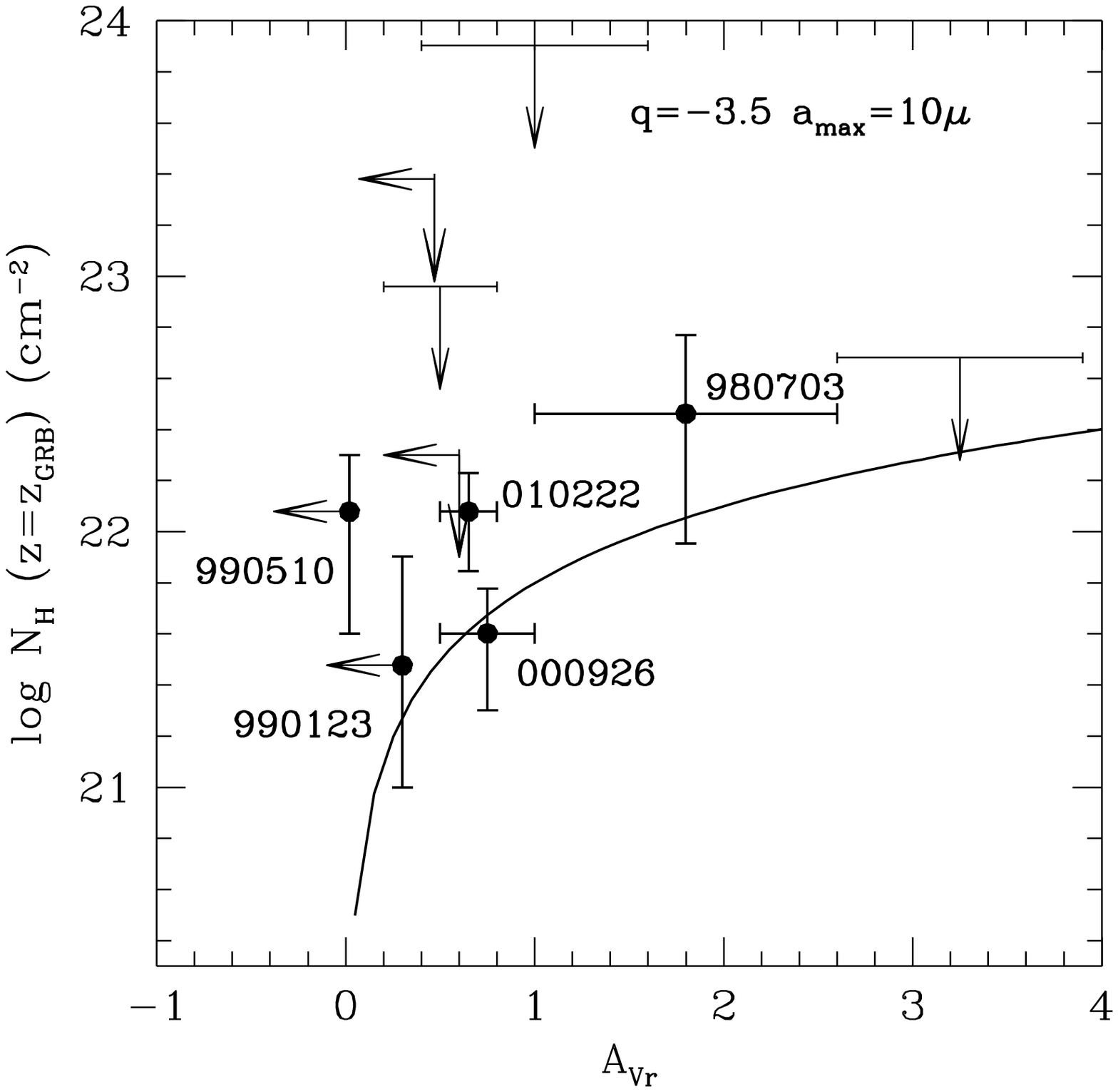}
\caption{
Best fit additional column densities against best fit visual extinction $A_{Vr}$ (both at the redshift of the GRB or at z=1 for the GRB with unknown redshifts) obtained using the Maiolino et al. (2001a) extinction curve (Q1). 
The $A_{Vr}$ are estimated using the $b_1$ ($\nu_o<\nu_c$) optical 
afterglow power law indices (see Tables 7 and 8). Error bars are 68\% confidence intervals and upper limits are 90\% confidence intervals. 
On the top-right of the plot are indicated the slope $q$ of the dust grain size distribution and the maximum size $a_{max}$ of the dust model used to derive the Q1 extinction curve. The solid line curve is the expected relationship between $N_H$ and $A_{Vr}$ for this models.\label{f12}}.
\end{figure}

\begin{figure}
\begin{center}
\plotone{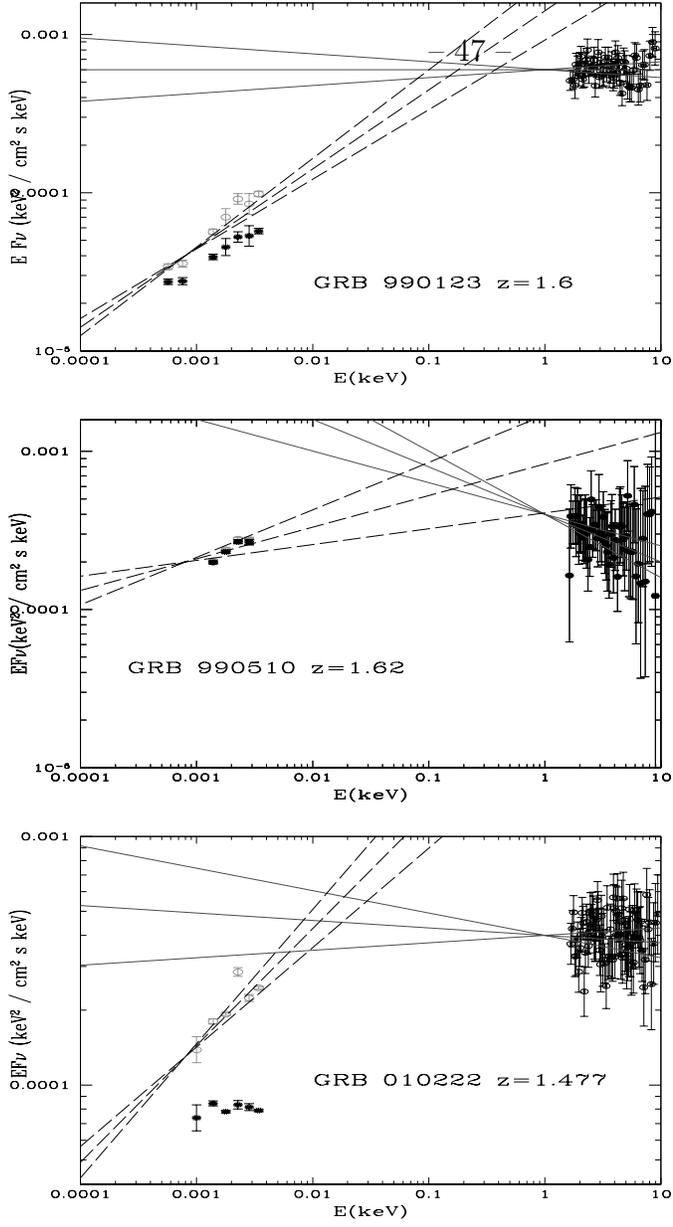}
\caption{
The $EF_E$ NIR to X-ray SED of GRB 990123, GRB 990510
and GRB 010222. Optical fluxes have been corrected for both the
Galactic and the extragalactic extinction. The latter was estimated 
using the best fit values obtained from the Q1 model for $\nu_o<\nu_c$. 
The normalization of the X-ray
spectra has been scaled to the same time of the optical-NIR
observations, using the published decay indices.
The solid lines are the best fit X-ray power laws and their 
90 \% upper and lower limits, normalized at 1 keV. 
The dashed lines are NIR to UV power laws, computed from the X-ray 
spectral indices assuming the fireball models and normalized at 1 
micron (observer frame).  \label{f13}} 
\end{center}
\end{figure}

{\bf GRB 971214}:
We refer to the Ramprakash et al. (1998) work that report K-band 
magnitudes and collect
previous V, R, I, J observations (Gorosabel et al. 1998, Odewahn et
al. 1998, Wijers \& Galama 1999, Ahn 2000) for this burst.
The Galactic extinction in the direction of this burst
is negligible (Halpern et al. 1998).  
A strong spectral curvature is evident in Figure \ref{f1} for this afterglow.
Wijers \& Galama (1999) interprets this feature as due to a 
genuine break in the
emission spectrum, expected in the blast wave model when the cooling
frequency is in the optical range (Galama \& Wijers 2001 exclude
for this reason the NIR data from their analysis, when trying to
determine the rest frame extinction fitting the SED with a simple power law).
On the other hand, Halpern et al. (1998) and Ramaprakash et al. (1998)
interpret the spectral curvature in terms of extinction by intervening
dust. They point out that a simple $A_\lambda \propto 1/\lambda$ dust
extinction law does not fit properly the data and that a more complex
extinction law should be considered.
In the following we consider both possibilities.

{\bf GRB 980329}:
We take the photometric data from the work of Reichart et al. (1999b)
which present an extended and complete summary of all the
observations of this burst (but see also Palazzi et al. 1998).
The Galactic reddening in the direction of GRB 980329 is estimated as 
E(B-V)=0.073 mag using the SFD maps.

{\bf GRB 980519}:
We take the photometric data from the work of Jaunsen et
al. (2001).
Halpern et al. (1999) report two different values for
the Galactic reddening: E(B-V)=0.267 using the SFD maps and
E(B-V)=0.348 using the 21 cm Stark et al. (1992) maps and assuming a
standard $N_H$ to $E(B-V)$ conversion (Savage \& Mathis 1979). Jaunsen
et al. (2001) find E(B-V)=0.24 using an alternative method, similar
to the SFD determination. We adopt here the latter value.

{\bf GRB 980703}:
The first observations of this afterglow were performed in the NIR
band some $\sim$30 hours after the GRB event. Later optical observations
are strongly affected by the host galaxy emission. We considered here
only the NIR points from the work of Vreeswijk et al. (1999) which summarize 
most previous observations (Zapatero Osorio et al. 1998, 
Rhoads et al. 1998, Henden et al. 1998,
Bloom et al. 1998a, Pedersen et al. 1998, Djorgovski et al. 1998a,
Sokolov et al. 1998b, Castro-Tirado et al. 1999, but also see Berger, 
Kuulkarni \& Frail 2001, Holland et al. 2001, Frail et al. 2003).  
The Galactic extinction in the direction of GRB 980703 is $A_V=0.19$,
using the SFD maps.

{\bf GRB 990123}:

The optical and infrared observations of this afterglow, 
(Bloom et al. 1999, Holland et al. 2000), are well
summarized by Galama et al. (1999) and by Kulkarni et al. (1999) and
therefore we took the photometric data from these two works.
The Galactic extinction in the direction of GRB 990123 is $A_V=0.053$,
using SFD maps.

{\bf GRB 990510}:
We took the B, V, R, I photometry from the work of Stanek et
al. (1999) (but also see Harrison et al. 1999, Beuerman et al. 1999 
and Holland et al. 2000), 
which gives a complete summary of the optical observation
on this burst.  These authors report magnitudes already corrected for
the Galactic reddening E(B-V)=0.203, estimated using the SFD maps.

{\bf GRB 010222}:
Masetti et al. (2001) present a complete summary of the
optical to NIR photometry (Lee et al. 2001,
Cowsik et al. 2001, Jha et al. 2001, 
also see Frail et al. 2002, Mirabal et al. 2002a, Bj\"ornsson
et al. 2002, Galama et al. 2003). 
A recent work from Frail et al. (2002) rules out that the
K-band emission is originated from the burst or its afterglow and
concludes that it is instead due to the emission of a high redshift
starburst galaxy. We therefore excluded the K band point in our analysis.
The galactic extinction in the direction of this burst 
is $A_V$=0.07, using the SFD maps.

\clearpage

\clearpage

\clearpage

\clearpage

\clearpage

\clearpage

\clearpage

\clearpage

\clearpage

\clearpage

\clearpage

\clearpage

\clearpage

\clearpage

\setcounter{page}{45}

\begin{center}
\begin{table}
\caption{\bf The BeppoSAX bright afterglow sample.\label{tbl-1}}
\begin{tabular}{lccccc}
\tableline\tableline
GRB    & $N_{H Gal}$\tablenotemark{a} & redshift & R.A.\tablenotemark{b}& Dec.\tablenotemark{b} & OT\tablenotemark{c}\\
       & $10^{20}$ cm$^{-2}$&         & h m s & $^\circ$ $'$ $''$ &\\
\hline
970228 & 16.0 & $0.6950\pm0.0003^{d}$	& 15 28 0.9  &  +19 37 02 & Y \\
970508 & 5.2  & $0.835\pm0.001^{e}$	& 06 53 46.7 &  +79 16 02 & Y \\
971214 & 1.6  & $3.418\pm0.010^{f}$	& 11 56 25   &  +65 13 11 & Y \\
980329 & 9.4  & -	     		& 07 02 36   &  +38 50 24 & Y \\
980519 & 1.7  & -     			& 23 22 22   &  +77 16 06 & Y \\
980703 & 5.8  & $0.9661\pm0.0001^{g}$	& 23 59 6.8  &  +08 35 45 & Y \\
990123 & 2.1  & $1.600\pm0.001^h$	& 15 25 30   &  +44 45 14 & Y \\
990510 & 9.4  & $1.619\pm0.002^i$ 	& 13 38 11.2 & -80 30 02  & Y \\
990704 & 3.1  & -     			& 12 19 27.3 & -03 50 22  & N  \\
000210 & 2.5  & $0.8463\pm0.0002^j$	& 01 59 09.8 & -40 39 15  & N  \\
000214 & 5.8  & 0.47$\pm0.06^k$       	& 18 54 27   & -66 27 30  & N \\
001109 & 4.2  & - 	& 18 30 07.8 	&  +55 17 56 & N \\
010222 & 1.6  & $1.4768\pm0.0002^l$	& 14 52 12.0 &  +43 01 02 & Y  \\
\tableline
\end{tabular}
\tablenotetext{a}{From Dickey \& Lockman (1990). Uncertainties on these values are 16\% at 1$\sigma$ confidence level (see \S 3.1)}
\tablenotetext{b}{From Narrow Field Instruments of BeppoSAX (J2000)}  
\tablenotetext{c}{Optical Transient}
\tablenotetext{d}{Bloom et al. 2001}
\tablenotetext{e}{Bloom et al. 1998b}
\tablenotetext{f}{Kulkarni et al. 1998}
\tablenotetext{g}{Djorgovski et al. 1998b}
\tablenotetext{h}{Kulkarni et al. 1999}
\tablenotetext{i}{Vreeswijk et al. 2000}
\tablenotetext{j}{Piro et al. 2002b}
\tablenotetext{k}{Antonelli et al. 2000 (from X-ray spectroscopy)}
\tablenotetext{l}{Mirabal et al. 2002c}
\end{table}
\end{center}

\clearpage

\begin{table}
\begin{scriptsize}
\begin{center}
\caption{\bf BeppoSAX LECS and MECS observations. \label{tbl-2}}
\begin{tabular}{lcccccccc}
\tableline\tableline
    &  \multicolumn{4}{c}{LECS 0.1-2.0keV} & \multicolumn{4}{c}{MECS 1.6-10.0keV}\\
GRB & Exposure$^a$ & r$^b$  & bkg$^c$   & source$^d$& Exposure$^a$ & r$^b$ & bkg$^c$   & source$^d$ \\
    & ks       & arcmin & $10^{-3}$ & $10^{-3}$ & ks       & arcmin & $10^{-3}$ & $10^{-3}$  \\
\hline
970228 & 5.5  & $6'$ & $3.5\pm0.1$ & $18.2\pm2.0$ & 14.3  & $3'$ & $1.5\pm0.1$	&$10.1\pm0.9$  \\
970508 & 17.2 & $6'$ & $3.8\pm0.1$ & $1.6\pm0.8$  & 28.2  & $3'$ & $5.5\pm0.1$	&$6.6\pm0.6$  \\
971214 & 45.5 & $6'$ & $4.1\pm0.1$ & $1.5\pm0.3$  & 101.1 & $3'$ & $3.5\pm0.1$	&$3.2\pm0.3$  \\
980329 & 24.5 & $3'$ & $1.2\pm0.1$ & $0.9\pm0.3$  & 63.8  & $3'$ & $3.5\pm0.1$	&$3.3\pm0.3$  \\
980519 & 22.7 & $4'$ & $2.1\pm0.1$ & $0.3\pm0.6$  & 77.8  & $3'$ & $3.5\pm0.1$	&$6.0\pm0.8$  \\
980703 & 16.3 & $4'$ &  $2.0\pm0.1$& $1.4\pm0.5$  & 39.1  & $4'$ & $3.4\pm0.1$	&$5.9\pm0.5$  \\
990123 & 15.1 & $4'$ & $2.1\pm0.1$ & $18.4\pm1.2$ & 45.2  & $3'$ & $3.5\pm0.1$	&$42.9\pm1.0$ \\
990510 & 31.2 & $4'$ &  $2.0\pm0.1$& $7.1\pm0.6$  & 67.9  & $3'$ & $3.5\pm0.1$	&$16.8\pm0.5$ \\
990704 & 13.4 & $4'$ & $2.1\pm0.1$ & $4.3\pm1.4$  & 37.0  & $3'$ & $3.5\pm0.1$ &$5.8\pm0.5$  \\
000210 & 15.7 & $6'$ & $4.6\pm0.1$ & $2.6\pm0.8$  & 44.4  & $3'$ & $3.5\pm0.1$ &$3.1\pm0.4$  \\
000214 & 14.8 & $4'$ & $2.1\pm0.1$ & $1.5\pm0.5$  & 50.8  & $3'$ & $3.5\pm0.1$	&$3.8\pm0.4$  \\
001109 & 19.2 & $4'$ & $2.1\pm0.1$ & $1.2\pm0.4$  & 85.4  & $3'$ & $3.6\pm0.1$	&$3.2\pm0.3$  \\
010222 & 50.3 & $4'$ & $2.1\pm0.1$ & $11.7\pm0.5$  & 88.4  & $3'$ &$3.5\pm0.1$ 	&$24.3\pm0.6$ \\
\tableline
\end{tabular}
\tablenotetext{a}{Net exposure time}
\tablenotetext{b}{Radius of source extraction region (see \S 3)}
\tablenotetext{c}{Background count rates}
\tablenotetext{d}{Net source count rates} 
\end{center}
\end{scriptsize}
\end{table}

\clearpage

\begin{table}
\begin{scriptsize}
\caption{\bf BeppoSAX NFI observations. \label{tbl-3}}
\begin{tabular}{lcccc}
\tableline\tableline
    & MECS     &  & &  LECS       \\
GRB & off-axis$^a$ &$(T_{start}-T_{GRB})^b$&$(T_{end}-T_{GRB})^b$ &RAW Coord.$^c$ \\
    & arcmin   & h  & h & Pixels \\
\hline
970228 & $0.4 $ & 8.0&17.0 & 118.4,155.0  \\
970508 & $0.4$  & 5.7&15.7 &112.9,116.6   \\
971214 & $1.9$  & 6.7&60.8 &132.6,123.5   \\
980329 & $1.6$  & 7.0&48.5 &134.2,126.9   \\
980519 & $1.7$  & 9.7&34.2 &135.6,135.8  \\
980703 & $1.7$  & 22.3&45.8&132.1,119.8   \\
990123 & $1.7$  & 5.8&32.2 &130.1,125.3 \\
990510 & $1.6$  & 8.0&52.3 &139.2,128.5 \\
990704 & $1.6$  & 8.0&29.9 &134.1,119.4  \\
000210 & $1.6$  & 7.2&38.9 &131.8,133.7   \\
000214 & $1.7$  & 12.0&40.9&138.0,119.2   \\
001109 & $1.7$  & 16.5&38.3&129.4,124.1   \\
010222 & $0.8$  & 8.0&65.0 &128.0,131.9\\
\tableline
\end{tabular}
\tablenotetext{a}{off-axis angle on the MECS detector}
\tablenotetext{b}{elapsed time of NFI first TOO from the time of the GRB detection}
\tablenotetext{c}{`raw' detector pixel coordinates of the source
at which the effective area has been computed.}
\end{scriptsize}
\end{table}

\clearpage

\begin{table}
\begin{scriptsize}
\caption{\bf The optical-near infrared magnitudes of the bright afterglow sample. \label{tbl-4}}
\begin{tabular}{lcccccccc}
\tableline\tableline
GRB(UT) &U & B  	& V 	 & R 	  & I  & J & H & K   \\
\hline
970228.12$^1$&-&- & $21.01\pm0.16$ & $20.91\pm0.29^{Rc}$ & $20.51\pm0.22^{Ic}$ &- &- &- \\
970508.90$^2$& $19.63\pm0.25$ & $20.59\pm0.08$ & $20.11\pm0.05$ & $19.86\pm0.05^{Rc}$ & $19.56\pm0.30^{Ic}$ &- &- &- \\
971214.97$^3$&- &    - 		&$22.62\pm0.11$	 & $22.07\pm0.10$ & $21.30\pm0.04$ & $20.45\pm0.25$ &- &$19.45\pm0.2$  \\
980329.15$^4$ &-&  -		&	-	 &  $23.30\pm0.20$&$20.89\pm0.30$ &$20.19\pm0.41$ &- &$19.03\pm0.5$ \\
980519.51$^5$ & $19.17\pm0.07$ & $19.74\pm0.03$ &$19.34\pm0.03$ & $18.92\pm0.03$ & $18.59\pm0.04$ &- &- &- \\
980703.18$^6$&- & 	-	& - &  $21.18\pm0.07$ & $20.54\pm0.61$ & -  & - &$17.55\pm0.47$\\
990123.41$^7$&$20.58\pm0.10$ & $21.26\pm0.37$& $20.93\pm0.19$ & $20.64\pm0.31$ & $20.29\pm0.10$ &- & $19.12\pm0.13$ & $18.29\pm0.07$  \\
990510.37$^8$&-&$19.51\pm0.11$ &$19.16\pm0.09$  & $18.87\pm0.08$&$18.53\pm0.07$&- &- &-\\
000926.99$^{10}$&$22.49\pm0.13$ & $22.6\pm0.12$& $22.03\pm0.12$ & $21.53\pm0.04$ & $21.01\pm0.05$ &$20.1\pm0.2$ & $18.8\pm0.1$ & $18.1\pm0.1$  \\
010222.31$^9$ & $20.23\pm0.03$ & $20.80\pm0.08$ & $20.43\pm0.10$ & $20.05\pm0.01$&$19.46\pm0.06$& $18.78\pm0.3$ & host gal? & host gal? \\
\tableline
\end{tabular}
$^1$ From Galama et al. (2000), $t_0$=UT 28.99 ($t_0$=extrapolation time, see \S 4. In this case this is at +20.9h from the gamma-ray event); $^2$Galama et al. (1998a), $t_0$=UT 10.98 (+49.8h); $^3$Ramaprakash et al. (1998), $t_0$=UT 15.51 (+12.9h); $^4$Reichart et al. (1999b), $t_0$=UT 29.99 (+19.9h); $^5$Jaunsen et al. (2001), $t_0$=UT 19.96 (+10.7h); $^6$Vreeswijk et al. (1999), $t_0$=UT 4.4 (+29.2h); $^7$Galama et al. (1999), $t_0$=UT 24.65 (+29.8h); $^8$Stanek et al. (1999), $t_0$=UT 11.26 (+21.4h); $^9$Masetti et al. (2001), $t_0$= UT 23.28 (+23.4h); $^{10}$Fynbo et al. (2001), $t_0$=UT 29.67 (+64.3h).
\end{scriptsize}
\end{table}

\clearpage

\begin{table}
\caption{\bf BeppoSAX LECS and MECS spectral fits in the 0.1-10.0 keV band. \label{tbl-5}}
\begin{tabular}{lcccccccc}
\tableline\tableline
GRB 	& LECS-MECS\tablenotemark{a} & $N_H$\tablenotemark{b} & z & $\alpha_X$\tablenotemark{c} & $\chi^2$(dof) & P(F)\tablenotemark{d} \\
    	& constant  & $10^{22}$ cm$^{-2}$ &  &   &  & \% \\
\hline
971214	& 0.7-1.4   & $80^{u.l.}$& 3.42 & $0.9^{+1.0}_{-0.6}$    & 34.12 (42) & 93.3 \\
        &           &  0 fixed               & 3.42 & $0.56\pm0.26$ & 36.99 (43) &  \\
        &           & $1.3^{u.l.}$           & 0    & $0.8^{+0.6}_{-0.4}$    & 34.79 (42) &  \\
\hline
990123	& 0.57-0.69 & $0.3^{+0.7}_{-0.2}$    & 1.60 & $1.0\pm0.1$    & 55.94 (66) & $>99.99$ \\
        &           &  0 fixed               & 1.60 & $0.9^{+0.08}_{-0.05}$  & 70.44 (67) &  \\
        &           & $0.04^{+0.07}_{-0.03}$ & 0    & $1.0\pm0.1$    & 55.73 (66) &  \\
\hline
990704	& 0.6-1.0   & $9^{u.l.}$    & 1    & $1.2^{+0.8}_{-0.5}$    & 17.36 (16) & 92.5 \\
        &           & 0 fixed       & 1    & $0.7\pm0.3$    	    & 21.29 (17) &  \\
        &           & 1.4$^{u.l.}$  & 0    & $1.0^{+0.3}_{-0.2}$    & 17.24 (16) &  \\
\hline
000210  & 0.9-2.2   & 8.0$^{u.l.}$           & 0.846& $1.5^{+0.9}_{-0.7}$    & 12.63 (13) & 89.9 \\
        &           & 0 fixed                & 0.846& $1.1^{+0.5}_{-0.6}$    & 15.65 (14) &      \\
        &	    & 1.8$^{u.l.}$           & 0    & $1.5^{+1.4}_{-0.7}$    & 12.77 (13) &      \\
\hline
001109	& 0.4-1.2   & $1.9^{+4.7}_{-1.8}$    & 1    & $1.4\pm0.6$    & 29.09 (27) & 97.0 \\
        &           &  0 fixed               & 1    & $1.1^{+0.3}_{-0.4}$    & 34.70 (28) &  \\
        &           & $0.4^{+0.9}_{-0.4}$ & 0    & $1.4^{+0.7}_{-0.6}$    & 29.10 (27) &  \\
\hline
010222  & 1.21-1.37 & $1.2^{+0.7}_{-0.6}$    & 1.477& $1.0\pm0.1$ & 99.80  (111) & $>99.99$\\
        &           &  0 fixed               & 1.477& $0.82^{+0.04}_{-0.08}$ & 146.11  (112)& \\
        &           & $0.15^{+0.08}_{-0.07}$ & 0    & $1.0\pm0.1$ & 104.61 (111) & \\
\tableline
\end{tabular}
\tablenotetext{ }{{\it u.l. = upper limit} at 90\% confidence level for two 
parameters of interest ($\Delta\chi^2=4.61$).}
\tablenotetext{a}{LECS-MECS normalization constant interval at 68\% confidence level}
\tablenotetext{b}{Equivalent hydrogen column density. Errors at 90\% confidence level for two parameters of interest ($\Delta\chi^2=4.61$).}
\tablenotetext{c}{Power law spectral index. Error at 90\% confidence level for two parameters of interest ($\Delta\chi^2=4.61$).}
\tablenotetext{d}{P(F) is the confidence level for the inclusion of additional column density, computed using the F test (Bevington \& Robinson 1992).}
\end{table}

\clearpage

\begin{table}
\caption{\bf BeppoSAX LECS and MECS spectral fits in the 0.4-10 keV band. \label{tbl-6}}
\begin{tabular}{lcccccccc}
\tableline\tableline
GRB 	& MECS-LECS & $N_H$  & z & $\alpha_X$ & $\chi^2$(dof) & P(F) \\
    	& constant  & $10^{22}$ cm$^{-2}$ & (fixed) &   &  &   \%\\
\hline
970228  & 0.6-1.0   &  $2^{u.l.}$            & 0.695 & $1.1\pm0.5$           & 6.90  (10) & 90.8\\
        &           & 0 fixed                & 0.695 & $0.8^{+0.3}_{-0.1}$   & 9.29  (11) & \\
        &           & $0.58^{u.l.}$          & 0     & $1.1\pm0.5$           & 7.00  (10) & \\
\hline
970508  & 0.4-1.0   & $9.1^{u.l.}$            & 0.835 & $0.8^{+0.4}_{-0.3}$   & 7.38  (12) & 90.7  \\
        &           & 0 fixed                & 0.835 & $0.5\pm0.3$           & 9.43  (13) & \\
        &           & $2.2^{u.l.}$           & 0     & $0.8^{+0.6}_{-0.4}$   & 7.43 (12) & \\
\hline
980329  & 0.6-0.9   & $4.8^{u.l.}$           & 1     & $1.1^{+0.7}_{-0.5}$   & 15.75 (17) & - \\
        &           & 0 fixed                & 1     & $1.1^{+0.5}_{-0.4}$   & 15.80 (18) & \\
 	& 	    & $1.0^{u.l.}$           & 0     & $1.2^{+0.6}_{-0.5}$   & 15.70  (17) & \\
\hline
980519  & 0.7-0.9   & $24^{u.l.}$            & 1     & $1.5^{+1.6}_{-1.0}$   & 15.95 (8)  & 50.3\\
        &           & 0 fixed                & 1     & $1.1^{+0.6}_{-0.5}$   & 16.96 (9)  & \\
        &           & $4.7^{u.l.}$           & 0     & $1.6^{+1.9}_{-1.1}$   & 15.91 (8)  & \\
\hline
980703  & 0.4-1.0   & $2.9^{+7.1}_{-2.7}$    & 0.966 & $1.8^{+0.8}_{-0.5}$   & 23.30 (20) & 97.5 \\
        &           & 0 fixed                & 0.966 & $1.3\pm0.3$   & 30.12 (21) & \\
        &           & $0.6^{+0.8}_{-0.4}$    & 0     & $1.7^{+1.3}_{-0.5}$   & 24.45 (20) & \\
\hline
990510  &  0.6-0.8  & $1.6^{+1.9}_{-1.3}$    & 1.62  & $1.3\pm0.2$   & 75.13 (65) & 98.3 \\
        &           & 0 fixed                & 1.62  & $1.16^{+0.08}_{-0.11}$& 83.11 (66) & \\
        &           & $0.16^{+0.24}_{-0.14}$ &  0    & $1.3\pm0.2$           & 76.43 (65) & \\
\hline
000214  &   0.3-0.7 & $0.1^{u.l.}$           & 0.47  & $2.0\pm0.6$           & 42.03 (20) & -  \\ 
        &           & 0 fixed                & 0.47  & $2.0\pm0.6$           & 41.96(21) &   \\ 
        & 	    & $0.08^{u.l.}$          & 0     & $2.0\pm0.6$           & 42.03(20)  &   \\
\tableline
\end{tabular}
\end{table}

\clearpage

\begin{table}
\caption{\bf Optical afterglow spectral fits. \label{tbl-7}}
\scriptsize
\begin{tabular}{lccccccc}
\tableline\tableline
GRB     & Models & \multicolumn{3}{c}{$\nu_o<\nu_c$}&\multicolumn{3}{c }{$\nu_o>\nu_c$} \\
\cline{3-5}\cline{6-8}
        &    	& $b_1$ &$A_{Vr}$ & $\chi^2/d.o.f.$ & $b_2$&  $A_{Vr}$ & $\chi^2/d.o.f.$ \\\cline{3-8}
\tableline
970228   & G  & $1.4\pm0.3$ &$<0.4$ & 0.95/1 & $0.9\pm0.3$ & $<0.2$ & 2.7/1 \\ 
z=0.695	 & SMC& $1.4\pm0.3$ &$<0.3$ & 0.95/1 & $0.9\pm0.3$ & $<0.1$ & 2.7/1 \\
$p=2.2\pm0.6$ & Q1 & $1.4\pm0.3$ &$<0.6$ & 0.96/1 & $0.9\pm0.3$ & $<0.3$ & 2.7/1 \\
	 & C  & $1.4\pm0.3$ &$<0.7$ & 0.90/1 & $0.9\pm0.3$ & $<0.4$ & 2.7/1 \\
\tableline
970508   & G  & $1.7\pm0.2$ & $0.27\pm0.16$ & 1.9/3 & $1.2\pm0.2$ &$<0.23$&4.3/3\\
z=0.835	 &SMC & $1.7\pm0.2$ & $0.23\pm0.17$ & 4.1/3 & $1.2\pm0.2$ &$<0.19$&5.0/3 \\
$p=1.6\pm0.4$ & Q1 & $1.7\pm0.2$ & $0.5\pm0.3$ & 2.2/3 & $1.2\pm0.2$ & $<0.4$& 4.4/3 \\
	 & C  & $1.7\pm0.2$ & $0.8\pm0.6$ & 4.8/3 & $1.2\pm0.2$ & $<0.7$& 7.5/3 \\
\tableline
971214   & G  & $1.6\pm0.5$ & $0.55\pm0.35$ & 4.3/1 & $1.1\pm0.5$ &$<0.6$& 2.5/1 \\
z=3.42	 &SMC & $1.6\pm0.5$ &$0.21\pm0.12$  & 0.7/1 &$1.1\pm0.5$ &$<0.24$&0.6/1\\
$p=1.8\pm1.0$ & Q1 & $1.6\pm0.5$ &$1.0\pm0.6$ & 2.9/1 & $1.1\pm0.5$ &$<1.2$& 1.8/1 \\
 	  & C & $1.6\pm0.5$ &$1.35\pm0.85$  & 1.1/1 & $1.1\pm0.5$ &$<1.7$& 0.8/1\\
%\tableline
&&&&&\\
971214$^{(*)}$& G  & $1.6\pm0.5$ &$<0.46$ & 10.2/3 & $1.1\pm0.5$ &$<0.5$&14.1/3 \\
  &SMC & $1.6\pm0.5$ &$0.2\pm0.1$ & 5.0/3&$1.1\pm0.5$&$<0.22$&11.0/3\\
	      & Q1 & $1.6\pm0.5$ &$0.76\pm0.74$& 8.9/3&$1.1\pm0.5$&$<0.1$&13.3/3\\
	      & C  & $1.6\pm0.5$ &$0.85\pm0.84$& 8.8/3&$1.1\pm0.5$&$<1.1$&13.7/3 \\
\tableline
980329  & G  & $1.40\pm0.35$ &$1.6\pm0.6$  & 20.8/2& $0.90\pm0.35$&$1.2\pm0.6$ & 21.2/2 \\
z=1     & SMC& $1.40\pm0.35$ &$1.45\pm0.55$& 17.2/2& $0.90\pm0.35$&$1.1\pm0.6$ & 18.4/2 \\
$p=2.2\pm0.7$& Q1 & $1.40\pm0.35$ &$3.25\pm0.65$& 17.8/2& $0.90\pm0.35$&$2.5\pm1.2$ & 18.9/2\\
	& C  & $1.40\pm0.35$ &$3.2\pm0.7$  &19.7/2& $0.90\pm0.35$&$2.35\pm1.15$ & 20.4/2\\
\tableline  
980519  & G   & $1.0\pm0.8$ &$<0.25$ & 29.3/3 & $0.5\pm0.8$ &$<0.03$ & 27.3/3\\
z=1	& SMC & $1.0\pm0.8$ &$<0.23$ & 16.3/3 & $0.5\pm0.8$ &$<0.02$ & 26.7/3\\
$p=3.0\pm1.6$& Q1  & $1.0\pm0.8$ &$<0.47$ & 38.0/3 & $0.5\pm0.8$ &$<0.05$ & 27.6/3\\
	& C   & $1.0\pm0.8$ &$<0.84$ & 28.1/3 & $0.5\pm0.8$ &$<0.02$ & 27.6/3\\
\tableline
980703  & G   & $0.7\pm0.4$ & $0.8\pm0.5$ & 3.6/1 & $0.2\pm0.4$ &$0.5\pm0.5$ &3.7/1 \\
z=0.966 & SMC & $0.7\pm0.4$ & $0.8\pm0.4$ & 7.2/1 & $0.2\pm0.4$&$0.4\pm0.3$ & 5.4/1 \\
$p=3.6\pm0.8$& Q1  & $0.7\pm0.4$ & $1.8\pm0.8$ & 7.4/1 & $0.2\pm0.4$&$0.9\pm0.8$ & 5.5/1 \\
	& C   & $0.7\pm0.4$ & $1.7\pm0.7$ & 5.0/1 & $0.2\pm0.4$&$0.9\pm0.8$ & 4.4/1\\*
\tableline
\end{tabular}
\tablecomments
{In the first column there are the burst name, the redshift and 
the electron spectral index value $p$ derived from the X-ray spectral index 
(Tab. 5-6) assuming that $\nu_X>\nu_c$. 
For GRB 000926 see \S 5.2. 
In the second column, there are the extinction models: 
G=Galactic, SMC=Small Magellanic Clouds,
Q1 is from dust model with large grain (q=3.5, $a_{max}=10\mu$), 
C=starburst galaxies extinction curve from Calzetti (1994). 
The $b_1$ and $b_2$ optical afterglow power law index 
are obtained from the electron spectral index value $p$ according 
to the standard fireball model for the case $\nu_o<\nu_c$ 
and $\nu_o>\nu_c$ (see \S 5.2).
Errors and upper limits are at $68\%$ confidence level.}

\tablenotetext{(*)}{We are here fitting also the J and K bands 
(see Appendix).}
\end{table}

\clearpage

\begin{table}
\caption{\bf Continue. \label{tbl-8}}
\begin{tabular}{lccccccccc}
\tableline\tableline
GRB     & Models & \multicolumn{3}{c}{$\nu_o<\nu_c$}&\multicolumn{3}{c }{$\nu_o>\nu_c$}\\
\cline{3-5}\cline{6-8}
        &    	& $b_1$ &$A_{Vr}$ & $\chi^2$ & $b_2$&  $A_{Vr}$ & $\chi^2$ \\\cline{3-8}
\tableline
990123   & G   & $1.50\pm0.06$ &$<0.14$ & 2.9/5 & $1.00\pm0.06$ &$<0.03$ &45.2/5\\
z=1.60	 & SMC & $1.50\pm0.06$ &$<0.06$ & 3.1/5  &$1.00\pm0.06$ &$<0.02$ &45.2/5\\
$p=2.0\pm0.1$ & Q1  & $1.50\pm0.06$ &$<0.3$ & 2.9/5 &$1.00\pm0.06$ &$<0.01$  & 45.2/5 \\
 	 & C   & $1.50\pm0.06$ &$<0.3$ & 2.6/5  &$1.00\pm0.06$ &$<0.01$  & 45.2/5 \\
\tableline
990510   & G   & $1.20\pm0.10$ &$<0.07$ & 22.6/2 & $0.7\pm0.1$&$<0.04$ & 62.9/2\\
z=1.62	 & SMC & $1.20\pm0.10$ &$<0.07$ & 22.6/2  & $0.7\pm0.1$&$<0.01$ & 64.1/2\\
$p=2.6\pm0.2$& Q1  & $1.20\pm0.10$ &$<0.02$ & 22.6/2  & $0.7\pm0.1$&$<0.01$ & 62.9/2\\
 	 & C   & $1.20\pm0.10$ &$<0.01$ & 22.6/2 & $0.7\pm0.1$&$<0.01$ & 62.9/2 \\
\tableline
000926 		& G  & $1.20\pm0.15$& $0.36\pm0.10$  & 26.9/6 & $0.70\pm0.15$ & $0.15\pm0.11$ & 22.6/6 \\
z=2.04 		& SMC& $1.20\pm0.15$& $0.22\pm0.10$ & 30.4/6 & $0.70\pm0.15$ & $0.10\pm0.07$ & 22.4/6 \\
$p=2.6\pm0.3$	& Q1 & $1.20\pm0.15$& $0.75\pm0.25$ & 23.1/6  &$0.70\pm0.15$ & $0.34\pm0.25$ & 21.1/6 \\
       		& C  & $1.20\pm0.15$& $1.34\pm0.29$ & 16.7/6 & $0.70\pm0.15$ & $0.41\pm0.30$ & 19.6/6 \\         
\tableline
010222   & G & $1.47\pm0.07$ &$0.31\pm0.11$ & 11.2/4& $0.97\pm0.07$&$<0.02$&7.7/4\\
z=1.477	 &SMC& $1.47\pm0.07$ &$0.10\pm0.03$ & 2.9/4 & $0.97\pm0.07$&$<0.1$ &7.6/4\\
$p=2.06\pm0.14$ & Q1& $1.47\pm0.07$ &$0.65\pm0.15$ & 10.8/4& $0.97\pm0.07$&$<0.06$& 7.7/4\\
	 & C & $1.47\pm0.07$ &$0.56\pm0.18$ & 2.4/4& $0.97\pm0.07$&$<0.04$&7.7/4\\
\tableline
\end{tabular}
\end{table}

\end{document}